\documentclass{sig-alternate-2013}

\newfont{\mycrnotice}{ptmr8t at 7pt}
\newfont{\myconfname}{ptmri8t at 7pt}
\let\crnotice\mycrnotice%
\let\confname\myconfname%

\permission{Permission to make digital or hard copies of all or part of this work for personal or classroom use is granted without fee provided that copies are not made or distributed for profit or commercial advantage and that copies bear this notice and the full citation on the first page. Copyrights for components of this work owned by others than ACM must be honored. Abstracting with credit is permitted. To copy otherwise, or republish, to post on servers or to redistribute to lists, requires prior specific permission and/or a fee. Request permissions from permissions@acm.org.}
\conferenceinfo{ICS'15,}{June 8--11, 2015, Newport Beach, CA, USA.\\
{\mycrnotice{Copyright is held by the owner/author(s). Publication rights licensed to ACM.}}}
\copyrightetc{ACM \the\acmcopyr}
\crdata{978-1-4503-3559-1/15/06\ ...\$15.00.\\
http://dx.doi.org/10.1145/2751205.2751209
}

\usepackage{amsmath}
\usepackage{array}
\usepackage{algorithm} 
\usepackage{algpseudocode}
\usepackage[caption=false]{subfig}
\usepackage{balance}
\usepackage{multirow}
\usepackage{hhline}
\usepackage{hyperref}
\usepackage{flushend}

\usepackage[
  pass,
]{geometry}

\begin{document}
%

\title{CSR5: An Efficient Storage Format for Cross-Platform Sparse Matrix-Vector Multiplication}

%
%
%
%
%

\numberofauthors{1} 
%
\author{
%
%
\alignauthor
Weifeng Liu, Brian Vinter\\
       \affaddr{Niels Bohr Institute, University of Copenhagen}\\
       \affaddr{Copenhagen, Denmark}\\
       \email{\{weifeng.liu, vinter\}@nbi.ku.dk}
}

\maketitle
\begin{abstract}
Sparse matrix-vector multiplication (SpMV) is a fundamental building block for numerous applications. In this paper, we propose CSR5 (Compressed Sparse Row 5), a new storage format, which offers high-throughput SpMV on various platforms including CPUs, GPUs and Xeon Phi. 
First, the CSR5 format is insensitive to the sparsity structure of the input matrix. Thus the single format can support an SpMV algorithm that is efficient both for regular matrices and for irregular matrices. Furthermore, we show that the overhead of the format conversion from the CSR to the CSR5 can be as low as the cost of a few SpMV operations.

We compare the CSR5-based SpMV algorithm\footnote{The source code of this work is downloadable at \url{https://github.com/bhSPARSE/Benchmark_SpMV_using_CSR5}.} with 11 state-of-the-art formats and algorithms on four mainstream processors using 14 regular and 10 irregular matrices as a benchmark suite. For the 14 regular matrices in the suite, we achieve comparable or better performance over the previous work. For the 10 irregular matrices, the CSR5 obtains average performance improvement of 17.6\%, 28.5\%, 173.0\% and 293.3\% (up to 213.3\%, 153.6\%, 405.1\% and 943.3\%) over the best existing work on dual-socket Intel CPUs, an nVidia GPU, an AMD GPU and an Intel Xeon Phi, respectively. For real-world applications such as a solver with only tens of iterations, the CSR5 format can be more practical because of its low-overhead for format conversion.
%


\end{abstract}

\category{G.1.3}{Numerical Linear Algebra}{Sparse, structured, and very large systems (direct and iterative methods)}
\category{G.4}{Mathematical Software}{Parallel and vector implementations}

\terms{Algorithms, Experimentation, Performance}

\keywords{Sparse Matrices, Storage Formats, SpMV, CSR, CSR5, CPU, GPU, Xeon Phi}

\section{Introduction} 

Over the past few decades, sparse matrix-vector multiplication (SpMV) has probably been the most studied sparse BLAS routine because of its importance in many scientific applications. The SpMV operation multiplies a sparse matrix $A$ of size $m\times n$ by a dense vector $x$ of size $n$ and obtains a dense vector $y$ of size $m$. Its na\"{\i}ve sequential implementation can be very simple, and can be easily parallelized by adding a few pragma directives for the compilers. But to accelerate large-scale computation, parallel SpMV is still required to be hand-optimized with specific data storage formats and algorithms~\cite{Ashari:Fast, Ashari:An, Baskaran:Optimizing, Bell:Implementing, Buluc:Reduced, Buluc:Parallel, Choi:Model, Deng:Taming, Garland:Sparse, Greathouse:Efficient, Kourtis:Exploiting, Li:SMAT, Li:GPU, Liu:Efficient, Su:clSpMV, Tang:Optimizing, Vuduc:OSKI, Williams:Optimization, Yan:yaSpMV}.

As a result, a conflict may emerge between the requirements of SpMV and other sparse matrix operations such as preconditioning operations~\cite{Li:GPU} and sparse matrix-matrix multiplication~\cite{Liu:An}. The reason is that those operations commonly require matrices stored in the basic formats such as the compressed sparse row (CSR). Therefore, when users construct a real-world application, they need to consider a cost of format conversion between the SpMV-oriented formats and the basic formats. Unfortunately, this conversion overhead may offset the benefits of using these specialized formats, in particular when only tens of iterations are needed in a solver.

The conversion cost is mainly from the expensive structure-dependent parameter tuning of a storage format. For example, some block-based formats require finding a good 2D block size~\cite{Buluc:Reduced, Buluc:Parallel, Choi:Model, Vuduc:OSKI, Vuduc:Automatic, Yan:yaSpMV}. Moreover, some hybrid formats~\cite{Bell:Implementing, Su:clSpMV} may need completely different partitioning parameters for distinct input matrices. 


To avoid the format conversion overhead, a few algorithms have concentrated on accelerating CSR-based SpMV with either row block methods~\cite{Ashari:Fast, Greathouse:Efficient} or segmented sum methods~\cite{Blelloch:Segmented, Garland:Sparse}. However, each of the two types of methods has its own drawbacks. As for the row block methods, despite their good performance for regular matrices, they may provide very low performance for irregular matrices due to unavoidable load imbalance. In contrast, the segmented sum methods can achieve near perfect load balance, but suffer from high overhead due to more global synchronizations and global memory accesses.
Furthermore, none of the above work can avoid an overhead from preprocessing, since certain auxiliary data for the basic CSR format have to be generated for better load balancing~\cite{Ashari:Fast, Greathouse:Efficient} or established primitives~\cite{Blelloch:Segmented, Garland:Sparse}.

Therefore, to be practical, an efficient format must satisfy two criteria: (1) it should limit format conversion cost by avoiding structure-dependent parameter tuning, and (2) it should support fast SpMV for both regular and irregular matrices.





To meet these two criteria, in this paper, we have designed CSR5 (Compressed Sparse Row 5)\footnote{The reason we call the storage format CSR5 is that it has five groups of data, instead of three in the classic CSR.}, a new format directly extending the classic CSR format. The CSR5 format leaves one of the three arrays of the CSR format unchanged, stores the other two arrays in an in-place tile-transposed order, and adds two groups of extra auxiliary information. The format conversion from the CSR to the CSR5 merely needs two tuning parameters: one is hardware-dependent and the other is sparsity-dependent (but structure-independent). Because the added two groups of information are usually much shorter than the original three in the CSR format, very limited extra space is required. Furthermore, the CSR5 format is SIMD-friendly and thus can be easily implemented on all mainstream processors with the SIMD units. Because of the structure-independence and the SIMD utilization, the CSR5-based SpMV algorithm can bring stable high throughput for both regular and irregular matrices. 



In this paper, we make the following contributions:

\begin{itemize}

  \item We propose CSR5, an efficient storage format with low conversion cost and high degree of parallelism.
  %

  \item We present a CSR5-based SpMV algorithm based on a redesigned low-overhead segmented sum algorithm.
  
  \item We implement the work on four mainstream devices: CPU, nVidia GPU, AMD GPU and Intel Xeon Phi.
  
  \item We evaluate the CSR5 format in both isolated SpMV tests and iteration-based scenarios. 
  
\end{itemize}

We compare the CSR5 with 11 state-of-the-art formats and algorithms on dual-socket Intel CPUs, an nVidia GPU, an AMD GPU and an Intel Xeon Phi. By using 14 regular and 10 irregular matrices as a benchmark suite, we show that the CSR5 obtains comparable or better performance over the previous work for the regular matrices, and can greatly outperform the prior work for the irregular matrices. As for the 10 irregular matrices, the CSR5 obtains average performance improvement of 17.6\%, 28.5\%, 173.0\% and 293.3\% (up to 213.3\%, 153.6\%, 405.1\% and 943.3\%) over the second best work on the four platforms, respectively. Moreover, for iteration-based real-world scenarios, the CSR5 format achieves higher speedups because of the fast format conversion. To the best of our knowledge, this is the first time that a single storage format can outperform state-of-the-art work on all four modern multicore and manycore processors.

\section{Preliminaries}

\subsection{The CSR Format}

\begin{figure}[h]
\begin{center}
\epsfig{file=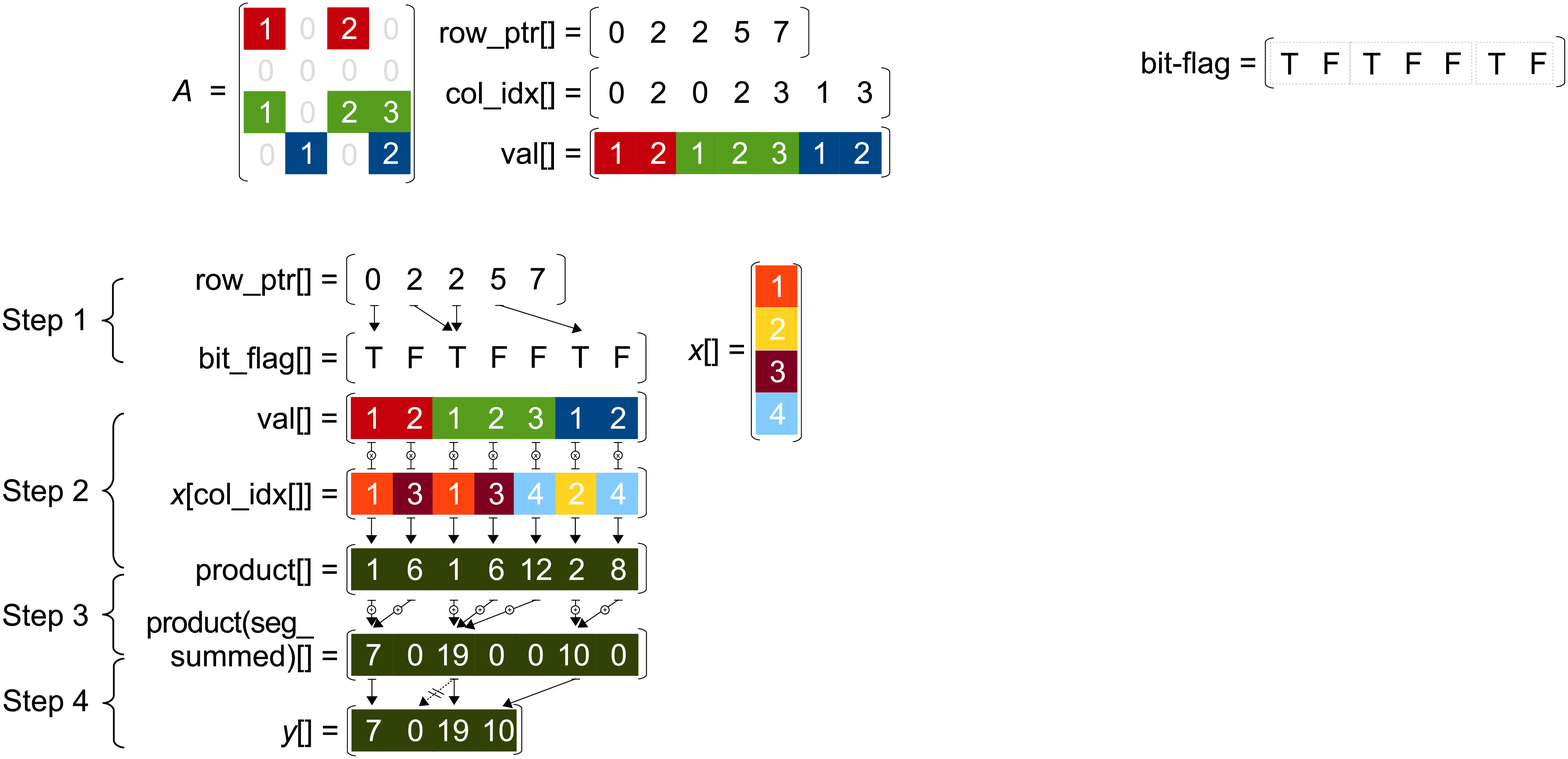, trim=2in 7.4in 13.1in 0.25in, clip=true, width=3in}
\end{center}
\vskip -16pt
\caption{A sparse matrix and its CSR format.}
\label{fig.csr}
\end{figure}

The CSR format for sparse matrices consists of three arrays: (1) \texttt{row\_ptr} array which saves the start and end pointers of the nonzeros of the rows. It has size $m+1$, where $m$ is the number of rows of the matrix, (2) \texttt{col\_idx} array of size $nnz$ stores column indices of the nonzeros, where $nnz$ is the number of nonzeros of the matrix, and (3) \texttt{val} array of size $nnz$ stores values of the nonzeros. Figure~\ref{fig.csr} shows an example.

\subsection{Parallel Algorithms for CSR-based SpMV}

\subsubsection{Row Block Methods}

In a given sparse matrix, rows are independent from each other. Therefore an SpMV operation can be parallelized on decomposed row blocks. A logical processing unit is responsible for a row block and stores dot product results of the matrix rows with the vector $x$ to corresponding locations in the result $y$. When the SIMD units of a physical processing unit are available, the SIMD reduction sum operation can be utilized for higher efficiency. These two methods are respectively known as the CSR-scalar and the CSR-vector algorithms, and have been implemented on CPUs~\cite{Williams:Optimization} and GPUs~\cite{Bell:Implementing, Su:clSpMV}. Algorithm~\ref{alg.scalar_spmv} shows a parallel CSR-scalar method.

\begin{algorithm}[h]
  \caption{SpMV using the CSR-scalar method.}\label{alg.scalar_spmv}
  \begin{algorithmic}[1]
    \For {$i=0$ \textbf{to} $m-1$ \textbf{in parallel}}
      \State \texttt{y[$i$]} $\gets 0$
      \For {$j=$ \texttt{row\_ptr[$i$]} \textbf{to} \texttt{row\_ptr[${i+1}$]$-1$}}
        \State \texttt{y[$i$]} $\gets$ \texttt{y[$i$]} $+$  \texttt{val[$j$]} $\times$ \texttt{x[col\_idx[$j$]]}
      \EndFor
    \EndFor
  \end{algorithmic}
\end{algorithm}

Despite the good parallelism, exploiting the scalability in modern multi-processors is not trivial for the row block methods. The performance problems mainly come from load imbalance for matrices which consist of rows with uneven lengths. Specifically, if one single row of a matrix is significantly longer than the other rows, only a single core can be fully used while the other cores in the same chip may be completely idle.
Although various strategies, such as data streaming~\cite{Dalton:CUSP, Greathouse:Efficient}, memory coalescing~\cite{Deng:Taming}, data reordering or reconstruction~\cite{Baskaran:Optimizing, Guo:Adaptive, Pinar:Improving}, static or dynamic binning~\cite{Ashari:Fast, Greathouse:Efficient} and Dynamic Parallelism~\cite{Ashari:Fast}, have been developed, none of those can fundamentally solve the problem of load imbalance, and thus provide relatively low SpMV performance for the CSR format.

\subsubsection{Segmented Sum Methods}

Blelloch et al.~\cite{Blelloch:Segmented} pointed out that the segmented sum may be more attractive for the CSR-based SpMV, since it is SIMD friendly and insensitive to the sparsity structure of the input matrix, thus overcoming the shortcomings of the row block methods.
%

Segmented sum (which is a special case of the backward segmented scan) performs a reduction sum operation for the entries in each segment in an array. A segment has its first entry flagged as \texttt{TRUE} and the other entries flagged as \texttt{FALSE}. Algorithm~\ref{alg.segsum} lists a serial segmented sum algorithm. 
Vectorized parallel segmented sum algorithms can be found in~\cite{Chatterjee:Scan, Dotsenko:Fast, Sengupta:Scan}.
%

\begin{algorithm}[h]
  \caption{Serial segmented sum operation.}\label{alg.segsum}
  \begin{algorithmic}[1]
    \Function{segmented\_sum}{\texttt{*in}, \texttt{*flag}}
      \State $length \gets$ \Call{sizeof}{\texttt{*in}}
      \For {$i=0$ \textbf{to} $length-1$ }
        \If {\texttt{flag[$i$]} $=$ \texttt{TRUE}}
          \State $j\gets i+1$
          \While{\texttt{flag[$j$]} $=$ \texttt{FALSE} \&\& $j < length$}
            \State \texttt{in[$i$]} $\gets$ \texttt{in[$i$]} $+$ \texttt{in[$j$]}
            \State $j\gets j+1$
          \EndWhile
        \Else
          \State \texttt{in[$i$]} $\gets 0$
        \EndIf
      \EndFor
    \EndFunction
  \end{algorithmic}
\end{algorithm}

\begin{figure}[!h]
\begin{center}
\epsfig{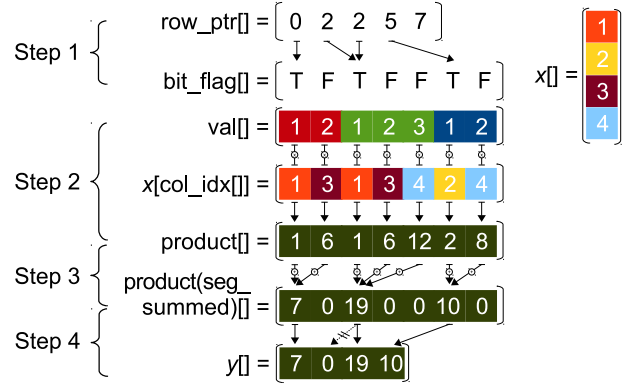}
\end{center}
\vskip -8pt
\caption{CSR-based SpMV using segmented sum.}
\label{fig.seg_sum_spmv}
\end{figure}

In the SpMV operation, the segmented sum treats each matrix row as a segment and calculates a partial sum for the entry-wise products generated in each row. 
The SpMV operation using the segmented sum methods consists of four steps: (1) generating an auxiliary \texttt{bit\_flag} array of size $nnz$ from the \texttt{row\_ptr} array. An entry in \texttt{bit\_flag} is flagged as \texttt{TRUE} if its location matches the first nonzero entry of a row, otherwise it is flagged as \texttt{FALSE}, (2) calculating all intermediate entries (i.e., entry-wise products) to an array of size $nnz$, (3) executing the parallel segmented sum for the array, and (4) collecting all partial sums to the result vector $y$ if a row is not empty. Algorithm~\ref{alg.segsum_spmv} lists the pseudocode. Figure~\ref{fig.seg_sum_spmv} illustrates an example using the matrix $A$ plotted in Figure 1. 

\begin{algorithm}[h]
  \caption{Segmented sum method CSR-based SpMV.}\label{alg.segsum_spmv}
  \begin{algorithmic}[1]
    \State \Call{malloc}{\texttt{*bit\_flag}, $nnz$}
    \State \Call{memset}{\texttt{*bit\_flag}, \texttt{FALSE}}
    \For {$i=0$ \textbf{to} $m-1$ \textbf{in parallel}} \Comment{Step 1}
      \State \texttt{bit\_flag[row\_ptr[$i$]] $\gets$ \texttt{TRUE}}
    \EndFor
    \State \Call{malloc}{\texttt{*product}, $nnz$} 
    \For {$j=0$ \textbf{to} $nnz-1$ \textbf{in parallel}} \Comment{Step 2}
      \State \texttt{product[$j$]} $\gets$ \texttt{val[$j$]} $\times$ \texttt{x[col\_idx[$j$]]}
    \EndFor
    \State \Call{segmented\_sum}{\texttt{*product}, \texttt{*bit\_flag}} \Comment{Step 3}
    \For {$k=0$ \textbf{to} $m-1$ \textbf{in parallel}} \Comment{Step 4}
      \If {\texttt{row\_ptr[$k$]} = \texttt{row\_ptr[$k+1$]}}
        \State \texttt{y[$k$]} $\gets$ 0
      \Else
        \State \texttt{y[$k$]} $\gets$ \texttt{product[row\_ptr[$k$]]}
      \EndIf
    \EndFor
    \State \Call{free}{\texttt{*bit\_flag}}
    \State \Call{free}{\texttt{*product}}
  \end{algorithmic}
\end{algorithm}

\begin{figure}[!h]
\begin{center}
\epsfig{file=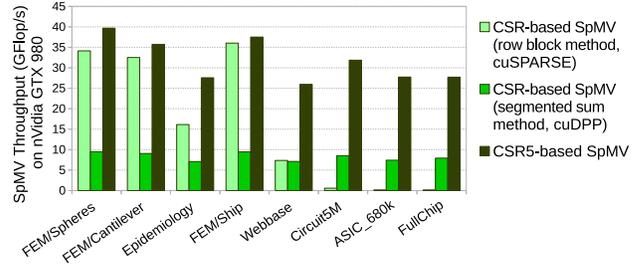, width=3.3in}
\end{center}
\vskip -8pt
\caption{Single precision SpMV performance.}
\label{fig.brief}
\end{figure}

We can see that once the heaviest workload, i.e., step 3, is parallelized through a fast segmented sum method described in~\cite{Chatterjee:Scan, Dotsenko:Fast, Sengupta:Scan}, nearly perfect load balance can be expected in all steps of Algorithm~\ref{alg.segsum_spmv}. However, in this context, the load balanced computation does not mean high performance. Figure~\ref{fig.brief} shows that the row block method in cuSPARSE v6.5 can significantly outperform the segmented sum method in cuDPP v2.2~\cite{Garland:Sparse, Sengupta:Scan}, while doing SpMV on relatively regular matrices (see Table~\ref{tab.suite} for the used benchmark suite). 

Why is this the case? We can see that the step 1 is a scatter operation and the step 4 is a gather operation, both from the row space of size $m$. This prevents the two steps from fusing with the steps 2 and 3 in the nonzero entry space of size $nnz$. In this case, more global synchronizations and global memory accesses may degrade the overall performance. Previous research~\cite{Bell:Implementing, Su:clSpMV} has found that the segmented sum may be more suitable for the COO (coordinate storage format) based SpMV, since the fully stored row index data can convert the steps 1 and 4 to the nonzero entry space: the \texttt{bit\_flag} array can be generated by comparison of neighbor row indices, and the partial sums in the \texttt{product} array can be directly saved to $y$ since their final locations are easily known from the row index array. 
Further, Yan et al.~\cite{Yan:yaSpMV} and Tang et al.~\cite{Tang:Optimizing} reported that some variants of the COO format can also benefit from the segmented sum. However, it is well known that accessing row indices in the COO pattern brings higher off-chip memory pressure, which is just what the CSR format tries to avoid. 

In the following, we will show that the CSR5-based SpMV can utilize both the segmented sum for load balance and the compressed row data for better load/store efficiency. In this way, the CSR5-based SpMV can obtain up to 4x speedup (see Figure~\ref{fig.brief}) over the CSR-based SpMV using the segmented sum primitive~\cite{Sengupta:Scan}.


\begin{figure*}[h!t]
\begin{center}
\epsfig{file=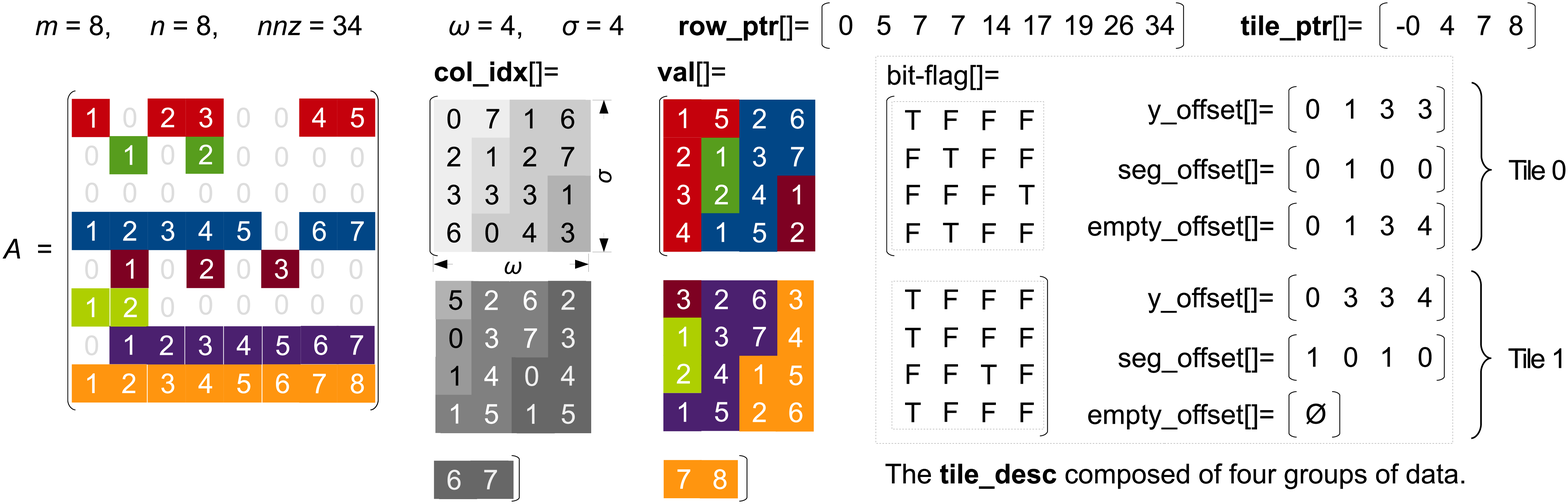, trim=0.1in 2.9in 3.1in 0.25in, clip=true, width=6.9in}
\end{center}
\vskip -16pt
\caption{The CSR5 storage format of a sparse matrix $A$ of size $8\times 8$. The five groups of information include \texttt{row\_ptr}, \texttt{tile\_ptr}, \texttt{col\_idx}, \texttt{val} and \texttt{tile\_desc}.}
\label{fig.csr5}
\end{figure*}

\section{The CSR5 Storage Format}

\subsection{Basic Data Layout}

To achieve near-optimal load balance for matrices with any sparsity structures, we first evenly partition all nonzero entries to multiple 2D tiles of the same size. Thus when executing parallel SpMV operation, a compute core can consume one or more 2D tiles, and each SIMD lane of the core can deal with one column of a tile. Then the main skeleton of the CSR5 format is simply a group of 2D tiles. The CSR5 format has two tuning parameters: $\omega$ and $\sigma$, where $\omega$ is a tile's width and $\sigma$ is its height. In fact, the CSR5 format \textit{only has} these two tuning parameters.

Further, we need extra information to efficiently compute SpMV. For each tile, we introduce a tile pointer \texttt{tile\_ptr} and a tile descriptor \texttt{tile\_desc}. Meanwhile, the three arrays, i.e., row pointer \texttt{row\_ptr}, column index \texttt{col\_idx} and value \texttt{val}, of the classic CSR format are directly integrated. The only difference is that the \texttt{col\_idx} data and the \texttt{val} data in each complete tile are in-place transposed (i.e., from row-major order to column-major order) for coalesced memory access from contiguous SIMD lanes. If the last entries of the matrix do not fill up a complete 2D tile (i.e., $nnz \mod (\omega \sigma) \neq 0$), they just remain unchanged and discard their \texttt{tile\_desc}. 

In Figure~\ref{fig.csr5}, an example matrix $A$ of size $8\times 8$ with $34$ nonzero entries is stored in the CSR5 format. When $\omega=4$ and $\sigma=4$, the matrix is divided into three tiles including two complete tiles of size 16 and one incomplete tile of size 2. The arrays \texttt{col\_idx} and \texttt{val} in the two complete tiles are stored in tile-level column-major order now. Moreover, only the first two tiles have \texttt{tile\_desc}, since they are complete.

\subsection{Auto-Tuned Parameters $\omega$ and $\sigma$}



Because the computational power of the modern multicore or manycore processors is mainly from the SIMD units, we design an auto-tuning strategy for high SIMD utilization.

First, the tile width $\omega$ is set to the size of the SIMD execution unit of the used processor. Then an SIMD unit can consume a 2D tile in $\sigma$ steps without any explicit synchronization, and the vector registers can be fully utilized. For the double precision SpMV, we always set $\omega=4$ for CPUs with 256-bit SIMD units, $\omega = 32$ for the nVidia GPUs, $\omega = 64$ for the AMD GPUs, and $\omega = 8$ for Intel Xeon Phi with 512-bit SIMD units. Therefore, $\omega$ can be automatically decided once the processor type used is known.

The other parameter $\sigma$ is decided by a slightly more complex process. For a given processor, we consider its on-chip memory strategy such as cache capacity and prefetching mechanism. If a 2D tile of size $\omega \times \sigma$ can empirically bring better performance than using the other sizes, the $\sigma$ is simply chosen. We found that the x86 processors fall into this category. For the double precision SpMV on CPUs and Xeon Phi, we always set $\sigma$ to 16 and 12, respectively. 

As for GPUs, the tile height $\sigma$ further depends on the sparsity of the matrix. Note that the ``sparsity'' is not equal to ``sparsity structure''. We define ``sparsity'' to be the average number of nonzero entries per row (or $nnz$/row for short). In contrast, ``sparsity structure'' is much more complex because it includes 2D space layout of all nonzero entries. 

On GPUs, we have several performance considerations on mapping the value $nnz$/row to $\sigma$. First, $\sigma$ should be large enough to expose more thread-level local work and to amortize a basic cost of the segmented sum algorithm. Second, it should not be too large since a larger tile potentially generates more partial sums (i.e., entries to store to $y$), which bring higher pressure to last level cache write. Moreover, for the matrices with large $nnz$/row, $\sigma$ may need to be small. The reason is that once the whole tile is located inside a matrix row (i.e., only one segment is in the tile), the segmented sum converts to a fast reduction sum. 

Therefore, for the $nnz$/row to $\sigma$ mapping on GPUs, we define three simple bounds: $r$, $s$ and $t$. The first bound $r$ is designed to prevent a too small $\sigma$. The second bound $s$ is used for preventing a too large $\sigma$. But when $nnz$/row is further larger than the third bound $t$, $\sigma$ is set to a small value $u$. Then we have 
\[ \sigma = \left\{ 
  \begin{array}{l l}
    r & \quad \text{if \;\; $nnz$/row $\leq r$}\\
    nnz$/row$ & \quad \text{if \;\; $r<$ $nnz$/row $\leq s$}\\
    s & \quad \text{if \;\; $s<$ $nnz$/row $\leq t$}\\
    u & \quad \text{if \;\; $t<$ $nnz$/row.}
  \end{array} \right.\]

The three bounds, $r$, $s$ and $t$, and the value $u$ are hardware-dependent, meaning that for a given processor, they can be fixed for use. For example, to execute double precision SpMV on nVidia Maxwell GPUs and AMD GCN GPUs, we always set $\texttt{<}r, s, t, u\texttt{>}=\texttt{<}4,32,256,4\texttt{>}$ and $\texttt{<}4,7,256,4\texttt{>}$, respectively. As for future processors with new architectures, we can obtain the four values through some simple benchmarks during initialization, and then use them for later runs. So the parameter $\sigma$ can be decided once the very basic information of a matrix and a low-level hardware are known. 



Therefore, we can see that the parameter tuning time becomes negligible because $\omega$ and $\sigma$ are easily obtained. This can save a great deal of preprocessing time. 



\subsection{Tile Pointer Information}

The added tile pointer information \texttt{tile\_ptr} stores the row index of the first matrix row in each tile, indicating the starting position for storing its partial sums to the vector $y$. By introducing \texttt{tile\_ptr}, each tile can find its own starting position, allowing tiles to execute in parallel.
The size of the \texttt{tile\_ptr} array is $p+1$, where $p = \lceil nnz / (\omega \sigma) \rceil$ is the number of tiles in the matrix. 
For the example in Figure~\ref{fig.csr5}, the first entry of Tile 1 is located in the $4th$ row of the matrix, and thus 4 is set as its tile pointer. To build the array, we binary search the index of the first nonzero entry of each tile on the \texttt{row\_ptr} array. Lines 1--4 in Algorithm~\ref{alg.tile_ptr} show this procedure.

Recall that an empty row has exactly the same row pointer information as its first non-empty right neighbor row (see the second row in the matrix $A$ in Figure~\ref{fig.csr}). 
Thus for the non-empty rows with an empty left neighbor, we need a specific process (which is similar to lines 12--16 in Algorithm~\ref{alg.segsum_spmv}) to store their partial sums to correct positions in $y$. To recognize whether the specific process is required, we give a hint to the other parts (i.e., tile descriptor data) of the CSR5 format and the CSR5-based SpMV algorithm. 
Here we set an entry in \texttt{tile\_ptr} to its negative value, if its corresponding tile includes any empty rows. Lines 5--12 in Algorithm~\ref{alg.tile_ptr} show this operation.

\begin{algorithm}[h]
  \caption{Generating \texttt{tile\_ptr}.}\label{alg.tile_ptr}
  \begin{algorithmic}[1]
    \For {$tid=0$ \textbf{to} $p$ \textbf{in parallel}} 
      \State \texttt{bnd} $\gets tid \times \omega \times \sigma$
      \State \texttt{tile\_ptr}[$tid$] $\gets$ \Call{binary\_search}{\texttt{*row\_ptr}, \texttt{bnd}}  $-1$
    \EndFor
    \For {$tid=0$ \textbf{to} $p-1$} 
      \For {$rid=$ \texttt{tile\_ptr}[$tid$] \textbf{to} \texttt{tile\_ptr}[$tid+1$]} 
        \If {\texttt{row\_ptr}[$rid$] $=$ \texttt{row\_ptr}[$rid+1$]}
          \State \texttt{tile\_ptr}[$tid$] $\gets$ \Call{negative}{\texttt{tile\_ptr}[$tid$]}
          \State \textbf{break}
        \EndIf
      \EndFor
    \EndFor
  \end{algorithmic}
\end{algorithm}

If the first tile has any empty rows, we need to store a $-0$ (negative zero) for it. To record $-0$, here we use unsigned 32- or 64-bit integer as data type of the \texttt{tile\_ptr} array. Therefore, we have 1 bit for explicitly storing the sign and 31 or 63 bits for an index. For example, in our design, tile pointer $-0$ is represented as a binary style `1000 ... 000', and tile pointer $0$ is stored as `0000 ... 000'. To the best of our knowledge, the index of 31 or 63 bits is completely compatible to most numerical libraries such as Intel MKL. Moreover, reference implementation of the recent high performance conjugate gradient (HPCG) benchmark~\cite{Dongarra:Toward} also uses 32-bit signed integer for problem dimension no more than $2^{31}$ and 64-bit signed integer for problem dimension larger than that. Thus it is safe to save 1 bit as the empty row hint and the other 31 or 63 bits as a `real' row index.

\subsection{Tile Descriptor Information}

Only having the tile pointer is not enough for a fast SpMV operation. For each tile, we also need four extra hints: (1) \texttt{bit\_flag} of size $\omega \times \sigma$, which indicates whether an entry is the first nonzero of a matrix row, (2) \texttt{y\_offset} of size $\omega$ used to further let each column know where the starting point to store its local partial sums is, (3) \texttt{seg\_offset} of size $\omega$ used to accelerate the local segmented sum inside a tile, and (4) \texttt{empty\_offset} of unfixed size (but no longer than $\omega \times \sigma$) constructed to help the partial sums to find correct locations in $y$ if the tile includes any empty rows. The tile descriptor \texttt{tile\_desc} is defined to denote a combination of the above four groups of data.

Generating \texttt{bit\_flag} is straightforward. The procedure is very similar to lines 3--5 in Algorithm~\ref{alg.segsum_spmv}. The main difference is that the bit flags are saved in column-major order, which matches the in-place transposed \texttt{col\_idx} and \texttt{val}. Additionally, the first entry of each tile's  \texttt{bit\_flag} is set to \texttt{TRUE} for sealing the first segment from the top and letting 2D tiles to be independent from each other.

The array \texttt{y\_offset} of size $\omega$ is used to help the columns in each tile knowing where the starting points to store their partial sums to $y$ are. In other words, each column has one entry in the array \texttt{y\_offset} as a starting point offset for all segments in the same column. We save a row index offset (i.e., relative row index) for each column in \texttt{y\_offset}. Thus for the $i$th column in the $tid$th tile, by calculating \texttt{tile\_ptr[$tid$]} $+$ \texttt{y\_offset[$i$]}, the column knows where its own starting position in $y$ is. Thus the columns can work in a high degree of parallelism without waiting for a synchronization. Generating \texttt{y\_offset} is simple: each column counts the number of \texttt{TRUE}s in its previous columns' \texttt{bit\_flag} array. Consider Tile 1 in Figure~\ref{fig.csr5} as an example: because there are 3 \texttt{TRUE}s in the 1st column, the 2nd column's corresponding value in \texttt{y\_offset} is 3. In addition, since there are in total 4 \texttt{TRUE}s in the 1st, 2nd and 3rd columns' \texttt{bit\_flag}, Tile 1's \texttt{y\_offset[$3$]}$=4$. Algorithm~\ref{alg.y_offset_n_seg_offset} lists how to generate \texttt{y\_offset} for a single 2D tile in an SIMD-friendly way.

\begin{algorithm}[h]
  \caption{Generating \texttt{y\_offset} and \texttt{seg\_offset}.}\label{alg.y_offset_n_seg_offset}
  \begin{algorithmic}[1]
    \State \Call{malloc}{\texttt{*tmp\_bit}, $\omega$}
    \State \Call{memset}{\texttt{*tmp\_bit}, \texttt{FALSE}}
    \For {$i=0$ \textbf{to} $\omega-1$ \textbf{in parallel}} 
      \State \texttt{y\_offset}[$i$] $\gets 0$
      \For {$j=0$ \textbf{to} $\sigma-1$} 
        \State \texttt{y\_offset}[$i$] $\gets$ \texttt{y\_offset}[$i$] $+$ \texttt{bit\_flag}[$i$][$j$]
        \State \texttt{tmp\_bit}[$i$] $\gets$ \texttt{tmp\_bit}[$i$] $\lor$ \texttt{bit\_flag}[$i$][$j$]
      \EndFor
      \State \texttt{seg\_offset}[$i$] $\gets$ $1-$ \texttt{tmp\_bit}[$i$]
    \EndFor
    \State \Call{exclusive\_prefix\_sum\_scan}{\texttt{*y\_offset}}
    \State \Call{segmented\_sum}{\texttt{*seg\_offset}, \texttt{*tmp\_bit}} 
    \State \Call{free}{\texttt{*tmp\_bit}}
  \end{algorithmic}
\end{algorithm}



The third array \texttt{seg\_offset} of size $\omega$ is used for accelerating a local segmented sum in the workload of each tile. The local segmented sum is an essential step that synchronizes partial sums in a 2D tile (imagine multiple columns in the tile come from the same matrix row). In the previous segmented sum (or segmented scan) method~\cite{Blelloch:Segmented, Chatterjee:Scan, Sengupta:Scan, Dotsenko:Fast}, the local segmented sum is complex and not efficient enough. Thus we prepare \texttt{seg\_offset} as an auxiliary array to facilitate implementation of segmented sum by way of the prefix-sum scan, which is a well optimized fundamental primitive for the SIMD units.

To generate \texttt{seg\_offset}, we let each column search its right neighbor columns and count the number of contiguous columns without any \texttt{TRUE}s in their \texttt{bit\_flag}. Using Tile 0 in Figure~\ref{fig.csr5} as an example, its 2nd column has one and only one right neighbor column (the 3rd column) without any \texttt{TRUE}s in its \texttt{bit\_flag}. Thus the 2nd column's \texttt{seg\_offset} value is 1. In contrast, because the other three columns (the 1st, 3rd and 4th) do not have any `all \texttt{FALSE}' right neighbors, their values in \texttt{seg\_offset} is 0. Algorithm~\ref{alg.y_offset_n_seg_offset} shows how to generate \texttt{seg\_offset} using an SIMD-friendly method.

\begin{figure}[h!t]
\begin{center}
\epsfig{file=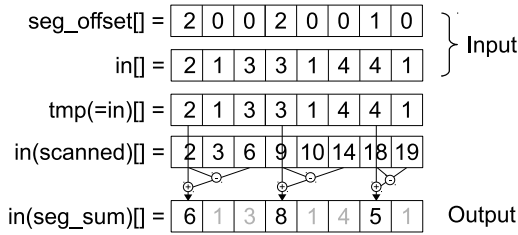, trim=0in 5in 15.8in 0.9in, clip=true, width=2.7in}
\end{center}
\vskip -16pt
\caption{An example of the fast segmented sum.}
\label{fig.fastsegsam}
\end{figure}

Algorithm~\ref{alg.fastsegsum} and Figure~\ref{fig.fastsegsam} show the fast segmented sum using \texttt{seg\_offset} and an inclusive prefix-sum scan. The principle of this operation is that the prefix-sum scan is essentially an increment operation. Once a segment knows the distance (i.e., offset) between its head and its tail, its partial sum can be deduced from its prefix-sum scan results. Therefore, the more complex segmented sum operation in~\cite{Blelloch:Segmented, Chatterjee:Scan, Sengupta:Scan, Dotsenko:Fast} can be converted to a faster prefix-sum scan operation (line 5) and a few arithmetic operations (lines 6--8).

\begin{algorithm}[h]
  \caption{Fast segmented sum using \texttt{seg\_offset}.}\label{alg.fastsegsum}
  \begin{algorithmic}[1]
    \Function{fast\_segmented\_sum}{$\texttt{*in}, \texttt{*seg\_offset}$}
      \State $length \gets$ \Call{sizeof}{\texttt{*in}}
      \State \Call{malloc}{\texttt{*tmp}, $length$}
      \State \Call{memcpy}{\texttt{*tmp}, \texttt{*in}}
      \State \Call{inclusive\_prefix\_sum\_scan}{\texttt{*in}}
      \For {$i=0$ \textbf{to} $length-1$ \textbf{in parallel}}
        \State \texttt{in[$i$]} $\gets$ \texttt{in[$i+$\texttt{seg\_offset[$i$]}]} $-$ \texttt{in[$i$]} $+$ \texttt{tmp[$i$]}
      \EndFor
      \State \Call{free}{\texttt{*tmp}}
    \EndFunction
  \end{algorithmic}
\end{algorithm}

The last array \texttt{empty\_offset} occurs when and only when a 2D tile includes any empty rows (i.e., its tile pointer is negative). Because an empty row of a matrix has the same row pointer with its rightmost non-empty neighbor row (recall the second row in the matrix $A$ in Figure 1), \texttt{y\_offset} will record an incorrect offset for it. We correct for this by storing correct offsets for segments within a tile. Thus the length of \texttt{empty\_offset} is the number of segments (i.e., the total number of \texttt{TRUE}s in \texttt{bit\_flag}) in a tile. For example, Tile 0 in Figure~\ref{fig.csr5} has 4 entries in its  \texttt{empty\_offset} since its \texttt{bit\_flag} includes 4 \texttt{TRUE}s. Algorithm~\ref{alg.empty_offset} lists the pseudocode that generates \texttt{empty\_offset} for a tile that contains at least one empty row.

\begin{algorithm}[h]
  \caption{Generating \texttt{empty\_offset} for the $tid$th tile.}\label{alg.empty_offset}
  \begin{algorithmic}[1]
    \State $length \gets$ \Call{reduction\_sum}{\texttt{*bit\_flag}}
    \State \Call{malloc}{\texttt{*empty\_offset}, $length$}
    \State $eid \gets 0$
    \For {$i=0$  \textbf{to} $\omega-1$} 
      \For {$j=0$  \textbf{to} $\sigma-1$} 
        \If {\texttt{bit\_flag[$i$][$j$]} $=$ \texttt{TRUE}}
          \State $ptr \gets tid \times \omega \times \sigma+i \times \sigma+j$
          \State \texttt{idx} $\gets$ \Call{binary\_search}{\texttt{*row\_ptr}, $ptr$}  $-1$ 
          \State \texttt{idx} $\gets$ \texttt{idx} $-$  \Call{remove\_sign}{\texttt{tile\_ptr[$tid$]}}
          \State \texttt{empty\_offset[$eid$]} $\gets$ \texttt{idx}
          \State $eid \gets eid + 1$
        \EndIf
      \EndFor
    \EndFor
  \end{algorithmic}
\end{algorithm}

\subsection{Storage Details}

To store the \texttt{tile\_desc} arrays in a space-efficient way, we find upper bounds to the entries and utilize the bit-field pattern. First, since entries in \texttt{y\_offset} store offset distances inside a 2D tile, they have an upper bound of $\omega \sigma$. So $\lceil \log_2 {(\omega \sigma)} \rceil$ bits are enough for each entry in \texttt{y\_offset}. For example, when $\omega=32$ and $\sigma=16$, 9 bits are enough for each entry. Second, since \texttt{seg\_offset} includes offsets less than $\omega$, $\lceil \log_2 {(\omega)} \rceil$ bits are enough for an entry in this array. For example, when $\omega=32$, 5 bits are enough for each entry. Third, \texttt{bit\_flag} stores $\sigma$ 1-bit flags for each column of a 2D tile. When $\sigma=16$, each column needs 16 bits. So 30 (i.e., $9+5+16$) bits are enough for each column in the example. Therefore, for a tile, the three arrays can be stored in a compact bit-field composed of $\omega$ 32-bit unsigned integers. If the above example matrix has 32-bit integer row index and 64-bit double precision values, only around $2\%$ extra space is required by the three newly added arrays. 

The size of \texttt{empty\_offset} depends on the number of groups of contiguous empty rows, since we only record one offset for the rightmost non-empty row with any number of empty rows as its left neighbors.

\subsection{The CSR5 for Other Matrix Operations}

Since we in-place transposed the CSR arrays \texttt{col\_idx} and \texttt{val}, a conversion from the CSR5 to the CSR is required for doing other sparse matrix operations using the CSR format. This conversion is simply removing \texttt{tile\_ptr} and \texttt{tile\_desc} and transposing \texttt{col\_idx} and \texttt{val} back to row-major order. Thus the conversion can be very fast. Further, since the CSR5 is a superset of the CSR, any entry accesses or slight changes can be done directly in the CSR5 format, without any need to convert it to the CSR format. Additionally, some applications such as finite element methods can directly assemble sparse matrices in the CSR5 format from data sources.

\section{The CSR5-based SpMV Algorithm}

Because all computations of the information (\texttt{tile\_ptr}, \texttt{tile\_desc}, \texttt{col\_idx} and \texttt{val}) of 2D tiles are independent of each other, they can execute concurrently. On GPUs, we assign a bunch of threads (i.e., warp in nVidia GPUs or wavefront in AMD GPUs) for each tile. On CPUs and Xeon Phi, we use OpenMP pragma for assigning the tiles to available x86 cores. Furthermore, the columns inside a tile are independent of each other as well. So we assign a thread on GPU cores or an SIMD lane on x86 cores to each column in a tile. 

While running the CSR5-based SpMV, each column in a tile can extract information from \texttt{bit\_flag} and label the segments in its local data to three colors: (1) \textit{red} means a sub-segment unsealed from its top, (2) \textit{green} means a completely sealed segment existed in the middle, and (3) \textit{blue} means a sub-segment unsealed from its bottom. There is an exception that if a column is unsealed both from its top and from its bottom, it is colored to \textit{red}.

\begin{figure}[h!t]
\begin{center}
\epsfig{file=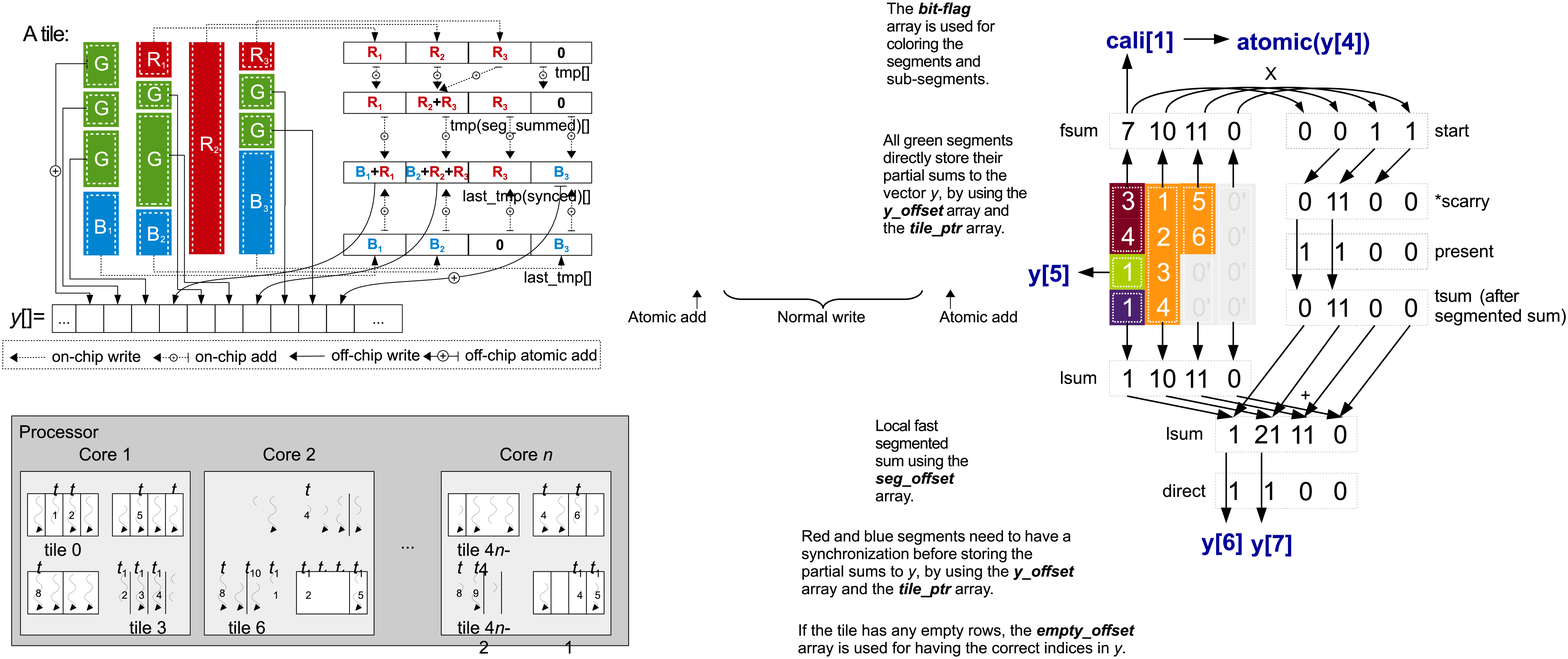, trim=1.6in 4.2in 13.6in 0.6in, clip=true, width=3.4in}
\end{center}
\vskip -18pt
\caption{The CSR5-based SpMV in a tile. Partial sums of the green segments are directly stored to $y$. The red and the blue sub-segments require an extra segmented sum before issuing off-chip write.}
\label{fig.csr5_spmv}
\end{figure}

Algorithm~\ref{alg.csr5_spmv} shows the pseudocode of the CSR5-based SpMV algorithm. Figure~\ref{fig.csr5_spmv} plots an example of this procedure. We can see that the green segments can directly save their partial sums to $y$ without any synchronization, since the indices can be calculated by using \texttt{tile\_ptr} and \texttt{y\_offset}. In contrast, the red and the blue sub-segments have to further add their partial sums together, since they are not complete segments. For example, the sub-segments B$_2$,  R$_2$ and R$_3$ in Figure~\ref{fig.csr5_spmv} have contributions to the same row, thus an addition is required. This addition operation needs the fast segmented sum shown in Algorithm~\ref{alg.fastsegsum} and Figure~\ref{fig.fastsegsam}. Furthermore, if a tile has any empty rows, the \texttt{empty\_offset} array is accessed to get correct global indices in $y$.  

\begin{algorithm}[h!]
  \caption{The CSR5-based SpMV for the $tid$th tile.}\label{alg.csr5_spmv}
  \begin{algorithmic}[1]
    \State \Call{malloc}{\texttt{*tmp}, $\omega$}
    \State \Call{memset}{\texttt{*tmp}, $0$}
    \State \Call{malloc}{\texttt{*last\_tmp}, $\omega$}
    \State \texttt{/*use \texttt{empty\_offset[\texttt{y\_offset[$i$]}]} instead of \texttt{y\_offset[$i$]} for a tile with any empty rows*/}
    \For {$i=0$ \textbf{to} $\omega-1$ \textbf{in parallel}} 
      \State \texttt{sum} $\gets 0$
      \For {$j=0$ \textbf{to} $\sigma-1$} 
        \State $ptr \gets tid \times \omega \times \sigma + j \times \omega + i$
        \State \texttt{sum} $\gets$ \texttt{sum} $+$ \texttt{val[$ptr$]} $\times$ \texttt{x[col\_idx[$ptr$]]}
        \State {\texttt{/*check bit\_flag[$i$][$j$]*/}}
        \If {\texttt{/*end of a red sub-segment*/}}
          \State \texttt{tmp[$i-1$]} $\gets$ \texttt{sum}
          \State \texttt{sum} $\gets 0$
        \ElsIf {\texttt{/*end of a green segment*/}}
          \State \texttt{y[\texttt{tile\_ptr[$tid$]} $+$ \texttt{y\_offset[$i$]}]} $\gets$ \texttt{sum}
          \State \texttt{y\_offset[$i$]} $\gets$ \texttt{y\_offset[$i$]} $+1$
          \State \texttt{sum} $\gets 0$
        \EndIf
      \EndFor
      \State \texttt{last\_tmp[$i$]} $\gets$ \texttt{sum} \texttt{//end of a blue sub-segment}
    \EndFor
    \State \Call{fast\_segmented\_sum}{\texttt{*tmp}, \texttt{*seg\_offset}} \Comment{Alg.~\ref{alg.fastsegsum}}
    \For {$i=0$ \textbf{to} $\omega-1$ \textbf{in parallel}} 
      \State \texttt{last\_tmp[$i$]} $\gets$ \texttt{last\_tmp[$i$]} $+$ \texttt{tmp[$i$]}
      \State \texttt{y[\texttt{tile\_ptr[$tid$]} $+$ \texttt{y\_offset[$i$]}]} $\gets$ \texttt{last\_tmp[$i$]}
    \EndFor
    \State \Call{free}{\texttt{*tmp}}
    \State \Call{free}{\texttt{*last\_tmp}}
  \end{algorithmic}
\end{algorithm}


Consider the synchronization among the tiles, since the same matrix row can be influenced by multiple 2D tiles running concurrently, the first and the last segments of a tile need to store to $y$ by atomic add (or a global auxiliary array used in device-level reduction, scan or segmented scan~\cite{Dotsenko:Fast, Sengupta:Scan}). In Figure~\ref{fig.csr5_spmv}, the atomic add operations are highlighted by arrow lines with plus signs.

For the last entries not in a complete tile (e.g., the last two nonzero entries of the matrix in Figure~\ref{fig.csr5}), we execute a conventional CSR-vector method after all of the complete 2D tiles have been consumed. Note that even though the last tile (i.e., the incomplete one) does not have \texttt{tile\_desc} arrays, it can extract a starting position from \texttt{tile\_ptr}.

In Algorithm~\ref{alg.csr5_spmv}, we can see that the main computation (lines 5--21) only contains very basic arithmetic and logic operations that can be easily programmed on all mainstream processors with SIMD units. As the most complex part in our algorithm, the fast segmented sum operation (line 22) only requires a prefix-sum scan, which has been well-studied and can be efficiently implemented by using CUDA, OpenCL or x86 SIMD intrinsics.

\begin{table*}[h!t]
\small 
\begin{center}
\begin{tabular}{|  >{\raggedright}m{5.5cm} |  >{\raggedright}m{11.38cm} |}
\hline\underline
    \centering \centering The testbeds & \centering The participating formats and algorithms \tabularnewline \hline
    \textbf{Dual-socket Intel Xeon E5-2667 v3} (Haswell, 2$\times$8 cores @ 3.2 GHz, 1.64 SP TFlops, 819.2 DP GFlops, 64 GB GDDR4, ECC-on, 2$\times$68.3 GB/s bandwidth). & (1) The CSR-based SpMV in Intel MKL 11.2 Update 1. \newline (2) BiCSB v1.2 using CSB~\cite{Buluc:Parallel} with bitmasked register block~\cite{Buluc:Reduced}. \newline (3) pOSKI v1.0.0~\cite{Byun:pOSKI} using OSKI v1.0.1h~\cite{Vuduc:OSKI, Vuduc:Automatic} kernels. \newline (4) The CSR5-based SpMV implemented by using OpenMP and AVX2 intrinsics.   \tabularnewline \hline
    \textbf{An nVidia GeForce GTX 980} (Maxwell GM204, 2048 CUDA cores @ 1.13 GHz, 4.61 SP TFlops, 144.1 DP GFlops, 4 GB GDDR5, 224 GB/s bandwidth, driver v344.16). & (1) The best CSR-based SpMV~\cite{Bell:Implementing} from cuSPARSE v6.5 and CUSP v0.4.0~\cite{Dalton:CUSP}.\newline (2) The best HYB~\cite{Bell:Implementing} from the above two libraries.\newline (3) BRC~\cite{Ashari:An} with texture cache enabled.\newline (4) ACSR~\cite{Ashari:Fast} with texture cache enabled. \newline (5) The CSR5-based SpMV implemented by using CUDA v6.5. \tabularnewline \hline
    \textbf{An AMD Radeon R9 290X} (GCN Hawaii, 2816 Radeon cores @ 1.05 GHz, 5.91 SP TFlops, 739.2 DP GFlops, 4 GB GDDR5, 345.6 GB/s bandwidth, driver v14.41). & (1) The CSR-vector method~\cite{Bell:Implementing} extracted from CUSP v0.4.0~\cite{Dalton:CUSP}.\newline (2) The CSR-Adaptive algorithm~\cite{Greathouse:Efficient} implemented in ViennaCL v1.6.2~\cite{Rupp:ViennaCL}. \newline (3) The CSR5-based SpMV implemented by using OpenCL v1.2. \tabularnewline \hline
    \textbf{An Intel Xeon Phi 5110p} (Knights Corner, 60 x86 cores @ 1.05 GHz, 2.02 SP TFlops, 1.01 DP TFlops, 8 GB GDDR5, ECC-on, 320 GB/s bandwidth, driver v3.4-1, $\mu$OS v2.6.38.8). &  (1) The CSR-based SpMV in Intel MKL 11.2 Update 1.\newline (2) The ESB~\cite{Liu:Efficient} with dynamic scheduling enabled.\newline (3) The CSR5-based SpMV implemented by using OpenMP and MIC-KNC intrinsics. \tabularnewline \hline
\end{tabular}
\end{center}
\vskip -16pt
\caption{The testbeds and participating formats and algorithms. }
\label{tab.testbeds}
\end{table*}

\section{Experimental Results}

\subsection{Experimental Setup}

We evaluate the CSR5-based SpMV and 11 state-of-the-art formats and algorithms on four mainstream platforms:  dual-socket Intel CPUs, an nVidia GPU, an AMD GPU and an Intel Xeon Phi. The platforms and participating approaches are shown in Table~\ref{tab.testbeds}.

Host of the two GPUs is a machine with AMD A10-7850K APU, dual-channel DDR3-1600 memory and 64-bit Ubuntu Linux v14.04 installed. Host of the Xeon Phi is a machine with Intel Xeon E5-2680 v2 CPU, quad-channel DDR3-1600 memory and 64-bit Red Hat Enterprise Linux v6.5 installed.
The two GPU platforms use the g++ compiler v4.8.2. The two Intel machines always set the Intel C/C++ complier 15.0.1 as default. 

Here we evaluate double precision SpMV. So cuDPP library~\cite{Garland:Sparse, Sengupta:Scan}, clSpMV~\cite{Su:clSpMV} and yaSpMV~\cite{Yan:yaSpMV} are not included since they only support single precision floating point as data type. Two recently published methods~\cite{Kreutzer:A, Tang:Optimizing} are not tested since the source code is not available to us yet. 

We use OpenCL profiling scheme for timing SpMV on the AMD platform and record wall-clock time on the other three platforms. For all participating formats and algorithms, we evaluate SpMV 10 times (each time contains 1000 runs and records the average) and report the best observed result. 

\subsection{Benchmark Suite}

In Table~\ref{tab.suite}, we list 24 sparse matrices as our benchmark suite for all platforms. The first 20 matrices have been widely adopted in previous SpMV research~\cite{Ashari:An, Bell:Implementing, Greathouse:Efficient, Liu:Efficient, Su:clSpMV, Williams:Optimization, Yan:yaSpMV}. The other 4 matrices are chosen since they have more diverse sparsity structures. All matrices except \textit{Dense} are downloadable at the University of Florida Sparse Matrix Collection~\cite{Davis:The}. 

To achieve a high degree of differentiation, we categorize the 24 matrices in Table~\ref{tab.suite} into two groups: (1) \textit{regular} group with the upper 14 matrices, (2) \textit{irregular} group with the lower 10 matrices. This classification is mainly based on the minimum, average and maximum lengths of the rows. Matrix \textit{dc2} is a representative of the group of irregular matrices. Its longest single row contains 114K nonzero entries, i.e., 15\% nonzero entries of the whole matrix with 117K rows. This sparsity pattern challenges the design of efficient storage format and SpMV algorithm.

\begin{table}[h!t]
\scriptsize
\begin{center}
\begin{tabular}{|  >{\raggedright}m{0.3cm} |  >{\raggedright}m{2cm} | >{\centering}m{1.45cm}  | >{\raggedleft}m{0.7cm}  | >{\centering}m{2cm} | }
\hline\underline
    \centering Id & Name &  $Dimensions$ & $nnz$ & $nnz$ per row \newline (min, avg, max) \tabularnewline \hline
    r1 & Dense &  2K$\times$2K & 4.0M & 2K, 2K, 2K  \tabularnewline 
    r2 & Protein & 36K$\times$36K & 4.3M & 18, 119, 204 \tabularnewline 
    r3 & FEM/Spheres & 83K$\times$83K & 6.0M & 1, 72, 81  \tabularnewline 
    r4 & FEM/Cantilever & 62K$\times$62K & 4.0M & 1, 64, 78  \tabularnewline 
    r5 & Wind Tunnel & 218K$\times$218K & 11.6M & 2, 53, 180  \tabularnewline 
    r6 & QCD & 49K$\times$49K & 1.9M & 39, 39, 39  \tabularnewline 
    r7 & Epidemiology & 526K$\times$526K & 2.1M & 2, 3, 4  \tabularnewline  
    r8 & FEM/Harbor & 47K$\times$47K & 2.4M & 4, 50, 145  \tabularnewline 
    r9 & FEM/Ship & 141K$\times$141K & 7.8M & 24, 55, 102  \tabularnewline 
    r10 & Economics & 207K$\times$207K & 1.3M & 1, 6, 44  \tabularnewline 
    r11 & FEM/Accelerator & 121K$\times$121K & 2.6M & 0, 21, 81  \tabularnewline 
    r12 & Circuit & 171K$\times$171K & 959K & 1, 5, 353  \tabularnewline 
    r13 & Ga41As41H72 & 268K$\times$268K & 18.5M & 18, 68, 702  \tabularnewline 
    r14 & Si41Ge41H72 & 186K$\times$186K & 15.0M & 13, 80, 662  \tabularnewline \hline 

    i1 & Webbase & 1M$\times$1M & 3.1M & 1, 3, 4.7K  \tabularnewline
    i2 & LP & 4K$\times$1.1M & 11.3M & 1, 2.6K, 56.2K  \tabularnewline 
    i3 & Circuit5M & 5.6M$\times$5.6M & 59.5M & 1, 10, 1.29M  \tabularnewline 
    i4 & eu-2005 & 863K$\times$863K & 19.2M & 0, 22, 6.9K  \tabularnewline 
    i5 & in-2004 & 1.4M$\times$1.4M & 16.9M & 0, 12, 7.8K  \tabularnewline 
    i6 & mip1 & 66K$\times$66K & 10.4M & 4, 155, 66.4K  \tabularnewline 
    i7 & ASIC\_680k & 683K$\times$683K & 3.9M & 1, 6, 395K  \tabularnewline 
    i8 & dc2 & 117K$\times$117K & 766K & 1, 7, 114K  \tabularnewline 
    i9 & FullChip & 2.9M$\times$2.9M & 26.6M & 1, 9, 2.3M  \tabularnewline 
    i10 & ins2 & 309K$\times$309K & 2.8M & 5, 9, 309K  \tabularnewline 
 \hline
\end{tabular}
\end{center}
\vskip -16pt
\caption{The benchmark suite. }
\label{tab.suite}
\end{table}

%


\subsection{Isolated SpMV Performance}

\begin{figure*}[h!t]
\captionsetup[subfigure]{labelformat=empty}
\centering
\subfloat[Legend]{\epsfig{file=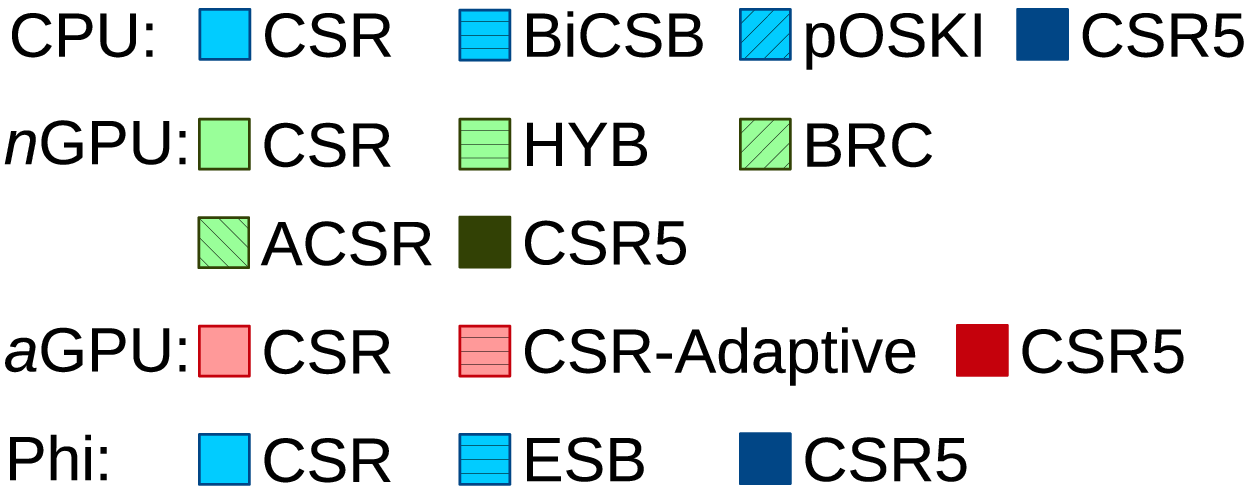, width=1.72in}}
\subfloat[(r1) Dense]{\epsfig{file=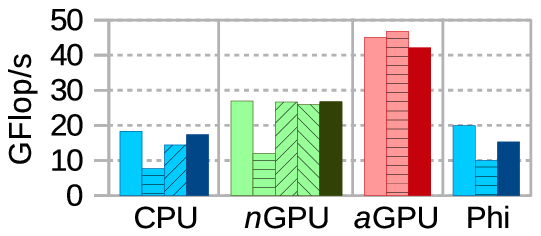, width=1.72in}}
\subfloat[(r2) Protein]{\epsfig{file=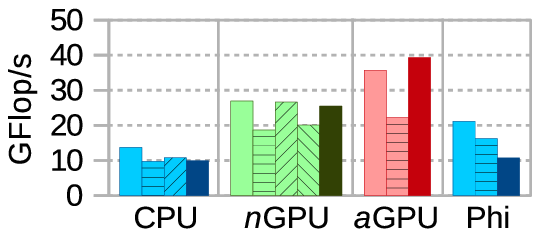, width=1.72in}}
\subfloat[(r3) FEM/Spheres]{\epsfig{file=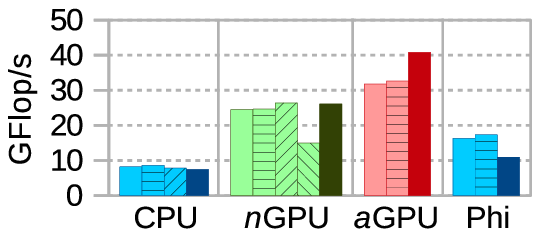, width=1.72in}}\qquad
\vskip -8pt
\subfloat[(r4) FEM/Cantilever]{\epsfig{file=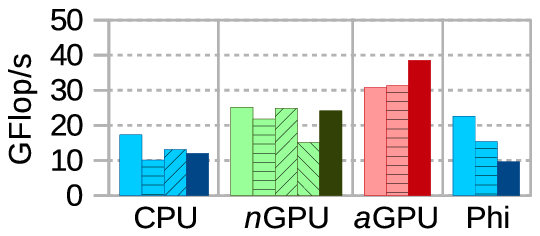, width=1.72in}}
\subfloat[(r5) Wind Tunnel]{\epsfig{file=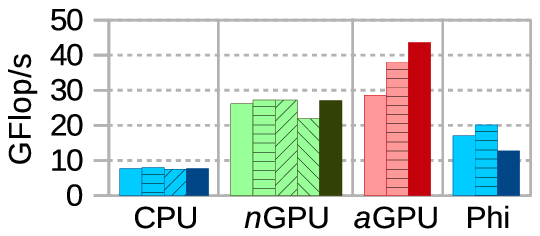, width=1.72in}}
\subfloat[(r6) QCD]{\epsfig{file=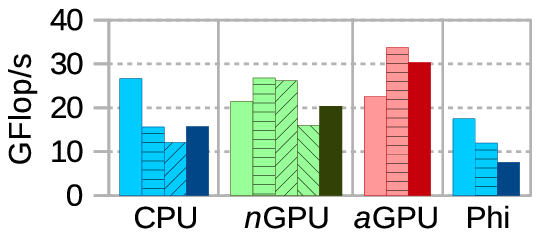, width=1.72in}}
\subfloat[(r7) Epidemiology]{\epsfig{file=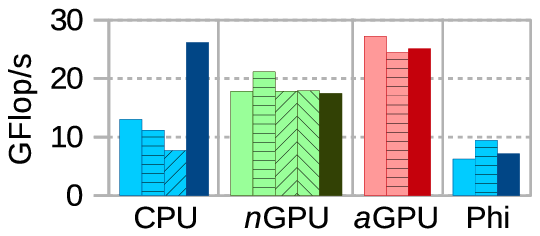, width=1.72in}}\qquad
\vskip -8pt
\subfloat[(r8) FEM/Harbor]{\epsfig{file=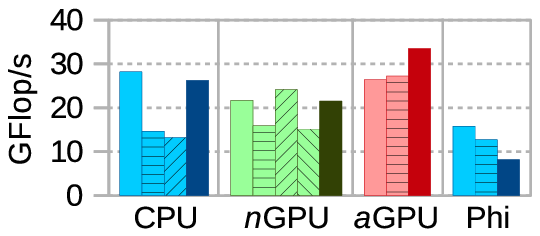, width=1.72in}}
\subfloat[(r9) FEM/Ship]{\epsfig{file=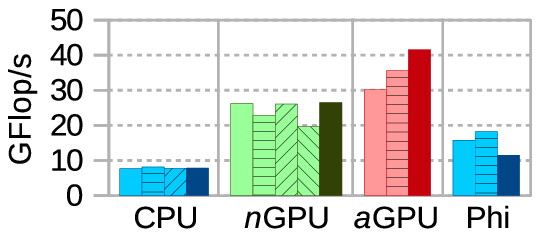, width=1.72in}}
\subfloat[(r10) Economics]{\epsfig{file=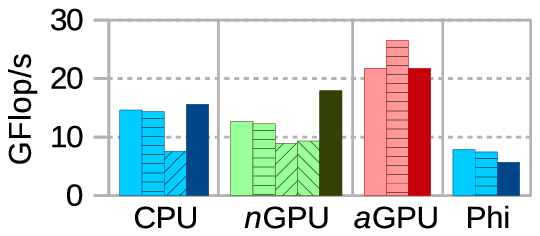, width=1.72in}}
\subfloat[(r11) FEM/Accelerator]{\epsfig{file=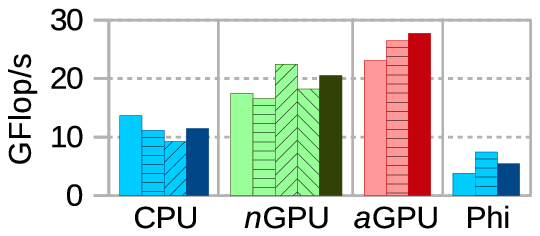, width=1.72in}}\qquad
\vskip -8pt
\subfloat[(r12) Circuit ]{\epsfig{file=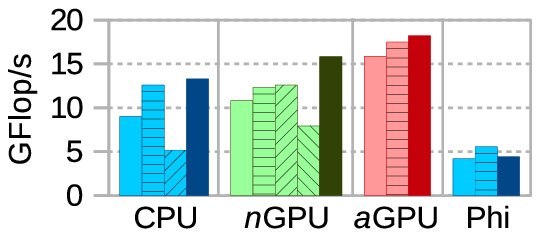, width=1.72in}}
\subfloat[(r13) Ga41As41H72]{\epsfig{file=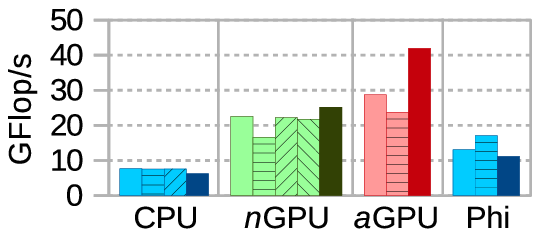, width=1.72in}}
\subfloat[(r14) Si41Ge41H72]{\epsfig{file=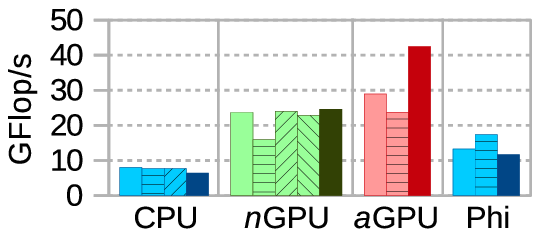, width=1.72in}}
\subfloat[(r1--r14) Harmonic mean]{\epsfig{file=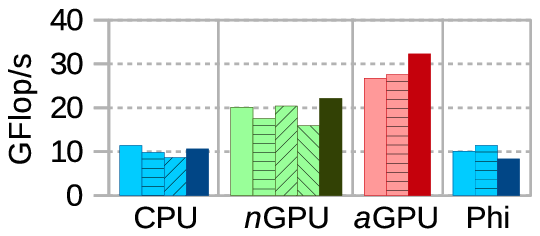, width=1.72in}}
\caption{The SpMV performance of the 14 regular matrices. (\textit{n}GPU$=$nVidia GPU, \textit{a}GPU$=$AMD GPU)}
\label{fig.benchr14}
\end{figure*}

Figure~\ref{fig.benchr14} shows double precision SpMV performance of the 14 regular matrices on the four platforms. We can see that, on average, all participating algorithms deliver comparable performance. \textit{On the CPU platform}, Intel MKL obtains the best performance on average and the other 3 methods behave similar. \textit{On the nVidia GPU}, the CSR5 delivers the highest throughput. The ACSR format is slower than the others, because its binning strategy leads to non-coalesced memory access. \textit{On the AMD GPU}, the CSR5 achieves the best performance. Although the dynamic assigning in the CSR-Adaptive method can obtain better scalability than the CSR-vector method, it still cannot achieve near perfect load balance. \textit{On the Xeon Phi}, the CSR5 is slower than Intel MKL and the ESB format. The main reason is that the current generation of Xeon Phi can only issue up to 4 relatively slow threads per core (i.e., up to $4\times 60$ threads in total on the used device), and thus the latency of gathering entries from vector $x$ becomes the main bottleneck. Then reordering or partitioning nonzero entries based on the column index for better cache locality behaves well in the ESB-based SpMV. However, in Section 5.6 we will show that this strategy leads to very high preprocessing cost.

\begin{figure*}[h!t]
\captionsetup[subfigure]{labelformat=empty}
\centering
\subfloat[Legend]{\epsfig{file=m-bench-legend.eps, width=1.72in}}
\subfloat[(i1) Webbase]{\epsfig{file=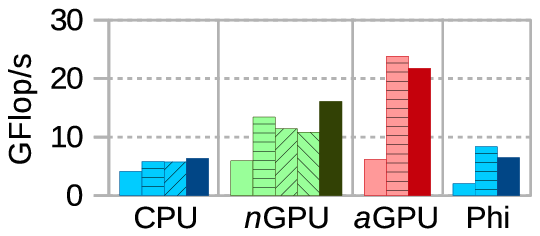, width=1.72in}}
\subfloat[(i2) LP]{\epsfig{file=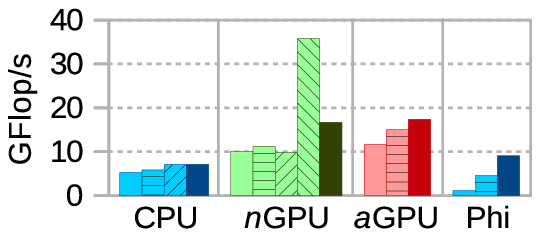, width=1.72in}}
\subfloat[(i3) Circuit5M]{\epsfig{file=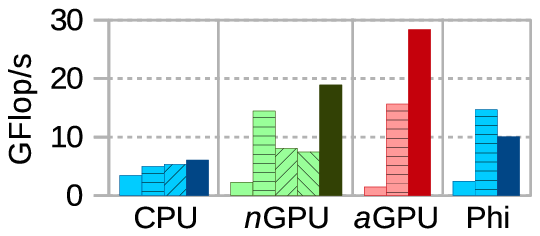, width=1.72in}}\qquad
\vskip -8pt
\subfloat[(i4) eu-2005]{\epsfig{file=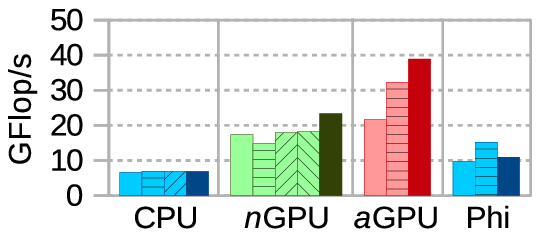, width=1.72in}}
\subfloat[(i5) in-2004]{\epsfig{file=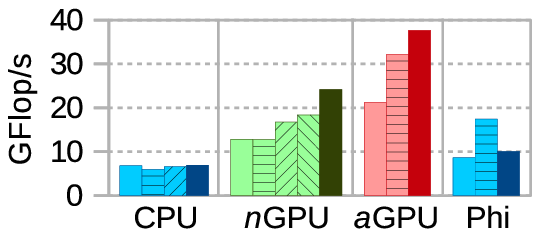, width=1.72in}}
\subfloat[(i6) mip1]{\epsfig{file=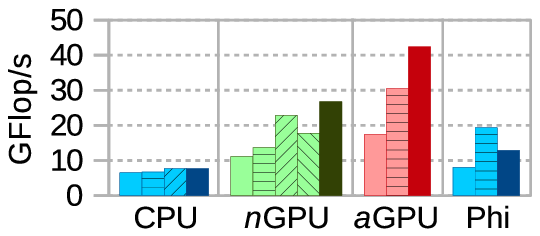, width=1.72in}}
\subfloat[(i7) ASIC\_680k]{\epsfig{file=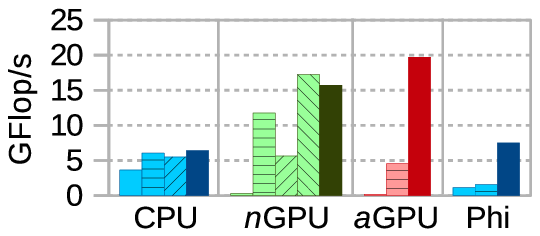, width=1.72in}}\qquad
\vskip -8pt
\subfloat[(i8) dc2]{\epsfig{file=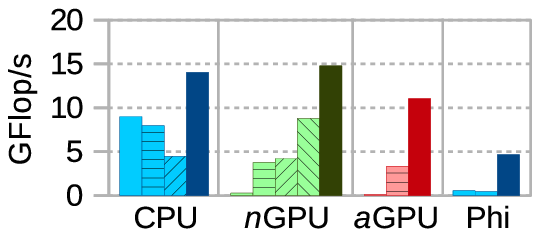, width=1.72in}}
\subfloat[(i9) FullChip]{\epsfig{file=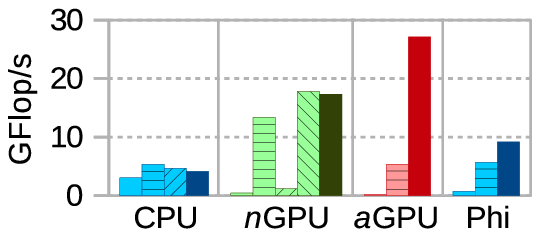, width=1.72in}}
\subfloat[(i10) ins2]{\epsfig{file=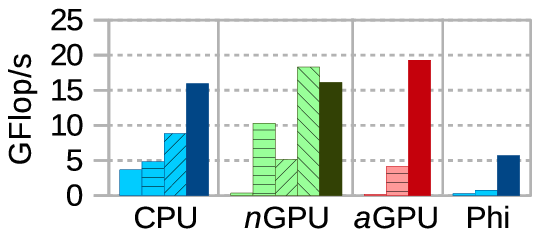, width=1.72in}}
\subfloat[(i1--i10) Harmonic mean]{\epsfig{file=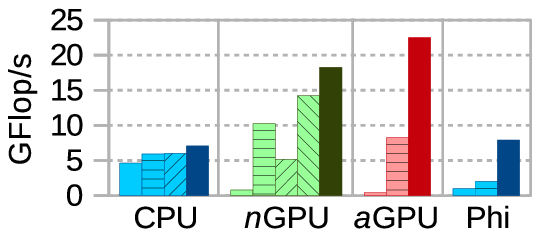, width=1.72in}}
\caption{The SpMV performance of the 10 irregular matrices. (\textit{n}GPU$=$nVidia GPU, \textit{a}GPU$=$AMD GPU)}
\label{fig.benchi10}
\end{figure*}

Figure~\ref{fig.benchi10} shows double precision SpMV performance of the 10 irregular matrices. We can see that the irregularity can dramatically impact SpMV throughput of some approaches. \textit{On the CPU platform}, the row block method based Intel MKL is now slower than the other methods. The CSR5 outperforms the others because of better SIMD efficiency from the AVX2 intrinsics. \textit{On the nVidia GPU}, the CSR5 brings the best performance because of the near perfect load balance. The other two irregularity-oriented formats, HYB and ACSR, behave well but still suffer from imbalanced work decomposition. Note that the ACSR format is based on Dynamic Parallelism, a technical feature only available on recently released nVidia GPUs. \textit{On the AMD GPU}, the CSR5 greatly outperforms the other two algorithms using the row block methods. Because the minimum work unit of the CSR-Adaptive method is one row, the method delivers degraded performance for matrices with very long rows\footnote{Note that we use an implementation of the CSR-Adaptive from the ViennaCL Library. The AMD's version of the CSR-Adaptive, which is not available to us yet, may have slightly different performance.}. \textit{On the Xeon Phi}, the CSR5 can greatly outperform the other two methods in particular when matrices are too irregular to expose cache locality of $x$ by the ESB format. Furthermore, since ESB is designed on top of the ELLPACK format, it cannot obtain the best performance for some irregular matrices.

Overall, the CSR5 achieves better performance (on the two GPU devices) or comparable performance (on the two x86 devices)  for the 14 regular matrices. For the 10 irregular matrices, compared to pOSKI, ACSR, CSR-Adaptive and ESB as the second best methods, the CSR5 obtains average performance gain of 17.6\%, 28.5\%, 173.0\% and 293.3\% (up to 213.3\%, 153.6\%, 405.1\% and 943.3\%), respectively.

\begin{table*}[h!t]
\small 
\centering
\begin{tabular}{|l|    r|r|r|r|r|    r|r|r|r|r|}
\hline
Benchmark & \multicolumn{5}{c|}{The 14 regular matrices} & \multicolumn{5}{c|}{The 10 irregular matrices} \tabularnewline \hline
\multirow{3}{*}{Metrics} & Preprocessing 	& \multicolumn{2}{c|}{Speedup} & \multicolumn{2}{c|}{Speedup}  & Preprocessing 	& \multicolumn{2}{c|}{Speedup}  		& \multicolumn{2}{c|}{Speedup}  		\\ \hhline{~~~~~~~~~~~}
& to SpMV & \multicolumn{2}{c|}{of \#iter.=50} & \multicolumn{2}{c|}{of \#iter.=500} & to SpMV  	& \multicolumn{2}{c|}{of \#iter.=50}  		& \multicolumn{2}{c|}{of \#iter.=500}  		  \\ \hhline{~~----~----}
			&  ratio	& avg  		& best  		& avg 		& best & ratio      & avg & best  & avg 		& best \\ \hline
CPU-BiCSB 	& 538.01x  & 0.06x  & 0.11x & 0.35x & 0.60x 				& 331.77x  & 0.13x  & 0.24x & 0.60x & 1.07x \\ 
CPU-pOSKI 	& 12.30x  & 0.43x  & 0.88x & 0.57x & 0.99x 				& 10.71x  & 0.62x  & 1.66x & 0.83x & 2.43x \\ 
CPU-CSR5 	& \textbf{6.14x}  & 0.52x  & 0.74x & 0.59x & 0.96x 				& \textbf{3.69x}  & 0.91x  & \textbf{2.37x} & \textbf{1.03x} & \textbf{2.93x} \\ \hline
\textit{n}GPU-HYB 	& 13.73x  & 0.73x  & 0.98x & 0.92x & 1.21x 			& 28.59x  & 1.86x  & 13.61x & 2.77x & 25.57x \\ 
\textit{n}GPU-BRC 	& 151.21x  & 0.26x  & 0.31x & 0.80x & 0.98x 			& 51.85x  & 1.17x  & 7.60x & 2.49x & 15.47x \\ 
\textit{n}GPU-ACSR 		& \textbf{1.10x} & 0.68x  & 0.93x & 0.72x & 1.03x 			& 3.04x & 5.05x  & 41.47x& 5.41x & 51.95x \\ 
\textit{n}GPU-CSR5 	& 3.06x  & \textbf{1.04x}  & \textbf{1.34x} & \textbf{1.10x} & \textbf{1.45x} 			& \textbf{1.99x}  & \textbf{6.43x}  & \textbf{48.37x} & \textbf{6.77x} & \textbf{52.31x} \\ \hline
\textit{a}GPU-CSR-Adaptive 	& \textbf{2.68x}  & 1.00x  & 1.33x & 1.07x & 1.48x 	& \textbf{1.16x}  & 3.02x  & 27.88x & 3.11x & 28.22x \\ 
\textit{a}GPU-CSR5 		& 4.99x & \textbf{1.04x}  & \textbf{1.39x} & \textbf{1.14x} & \textbf{1.51x} 			& 3.10x & \textbf{5.72x}  & \textbf{135.32x} & \textbf{6.04x} & \textbf{141.94x} \\ \hline
Phi-ESB 	& 922.47x  & 0.05x  & 0.15x & 0.33x & 0.88x 					& 222.19x  & 0.27x  & 1.15x & 1.30x & 2.96x \\ 
Phi-CSR5 	& \textbf{11.52x}  & 0.54x  & \textbf{1.14x} & 0.65x & \textbf{1.39x} 				& \textbf{9.45x}  & \textbf{3.43x}  & \textbf{18.48x} & \textbf{4.10x} & \textbf{21.18x} \\ \hline
\end{tabular}
\caption{Preprocessing cost and its impact on the iteration-based scenarios.}
\label{tab.iter}
\end{table*}

\subsection{Effects of Auto-Tuning}

\begin{figure}[h!t]
\centering
\subfloat[The nVidia GTX 980 GPU.]{\epsfig{file=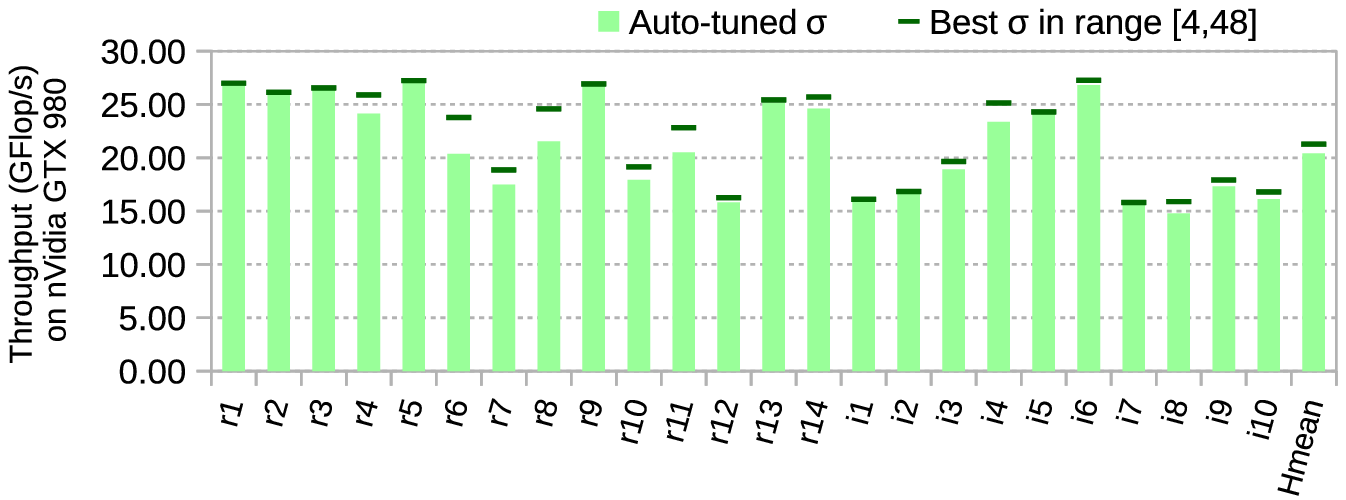, width=3.3in}}\qquad
\subfloat[The AMD R9-290X GPU.]{\epsfig{file=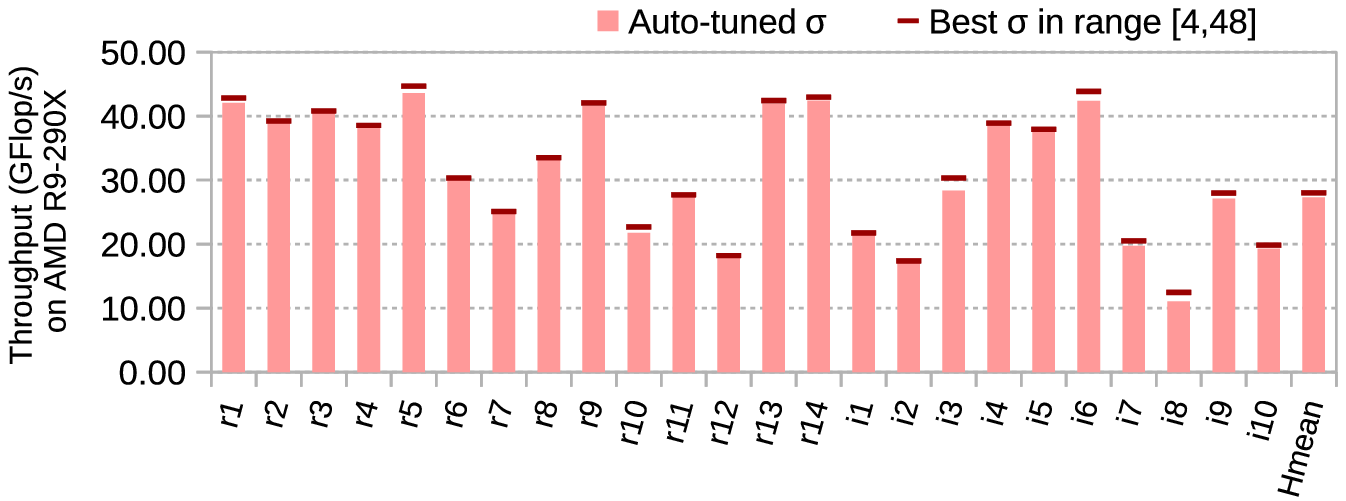, width=3.3in}}\qquad
\caption{Auto-tuning effects on the two GPUs.}
\label{fig.autotuning}
\end{figure}

In section 3.2, we discussed a simple auto-tuning scheme for the parameter $\sigma$ on GPUs. Figure~\ref{fig.autotuning} shows its effects (the x axis is the matrix \textit{id}s). We can see that compared to the best performance chosen from a range of $\sigma=$ 4 to 48, the auto-tuned $\sigma$ does not have obvious performance loss. On the nVidia GPU, the performance loss is on average -4.2\%. On the AMD GPU, the value is on average -2.5\%. 

\subsection{Format Conversion Cost}

\begin{figure}[h!t]
\captionsetup[subfigure]{labelformat=empty}
\centering
\subfloat[]{\epsfig{file=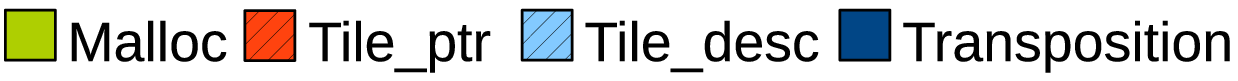, trim=0in 1.9in 0in 0in, clip=true, width=2.2in}} \qquad
\vskip -26pt
\subfloat[(a) The CPU.]{\epsfig{file=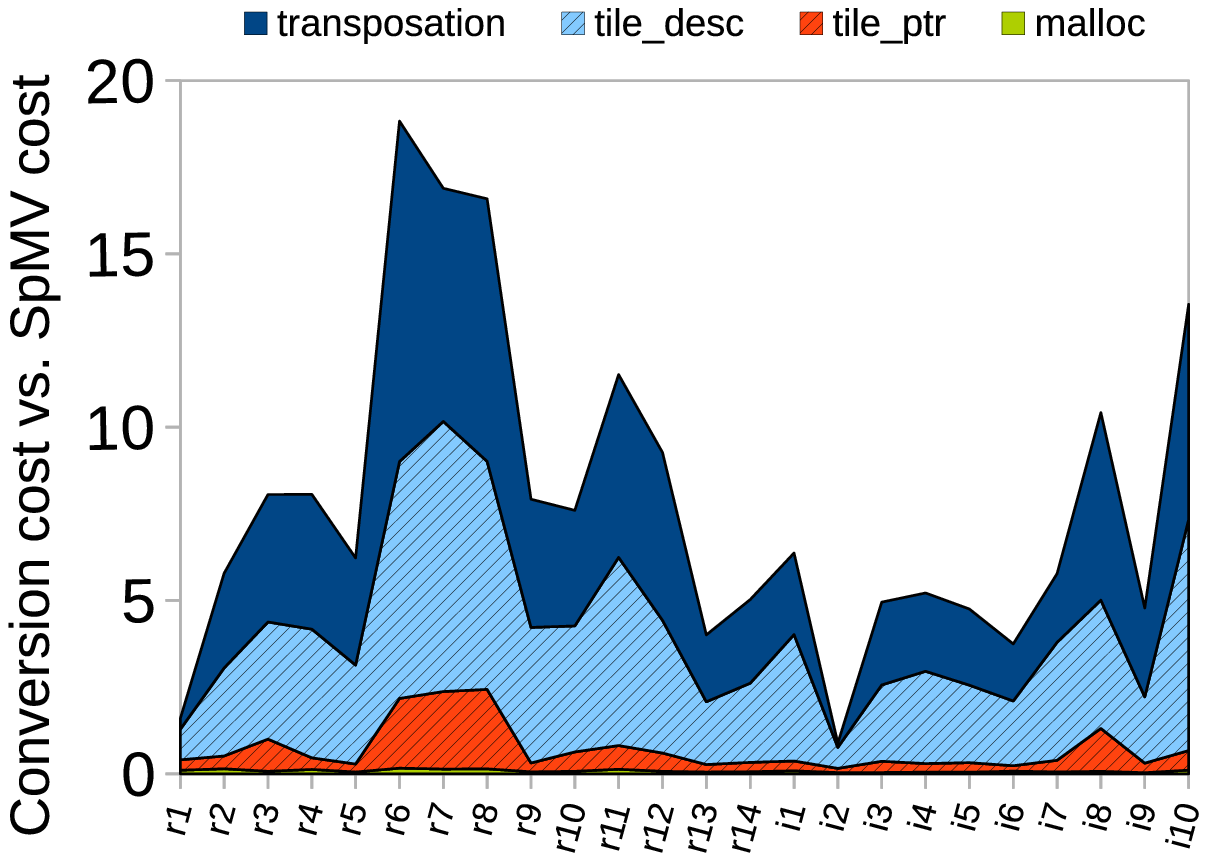, trim=0in 0in 0in 0.3in, clip=true, width=1.6in}}
\subfloat[(b) The nVidia GPU.]{\epsfig{file=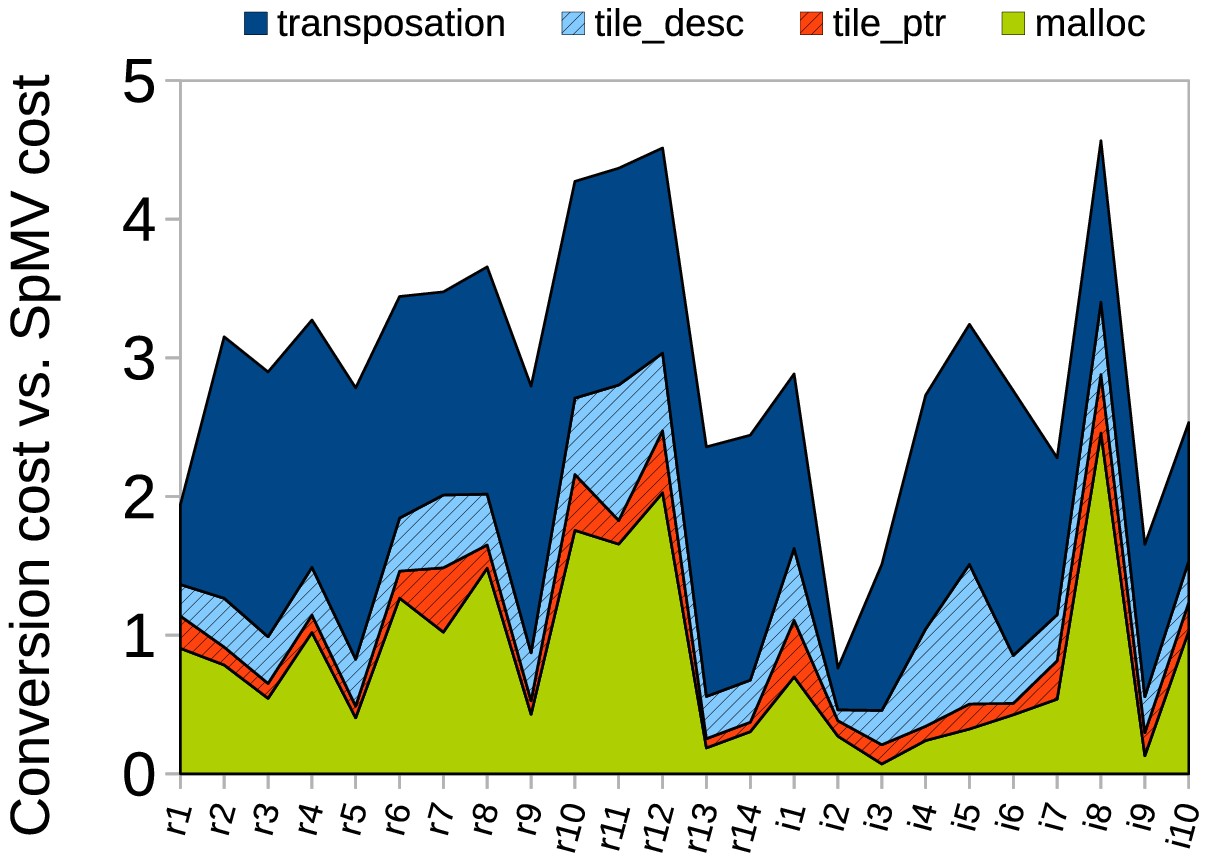, trim=0in 0in 0in 0.3in, clip=true, width=1.6in}} \qquad
\subfloat[(c) The AMD GPU.]{\epsfig{file=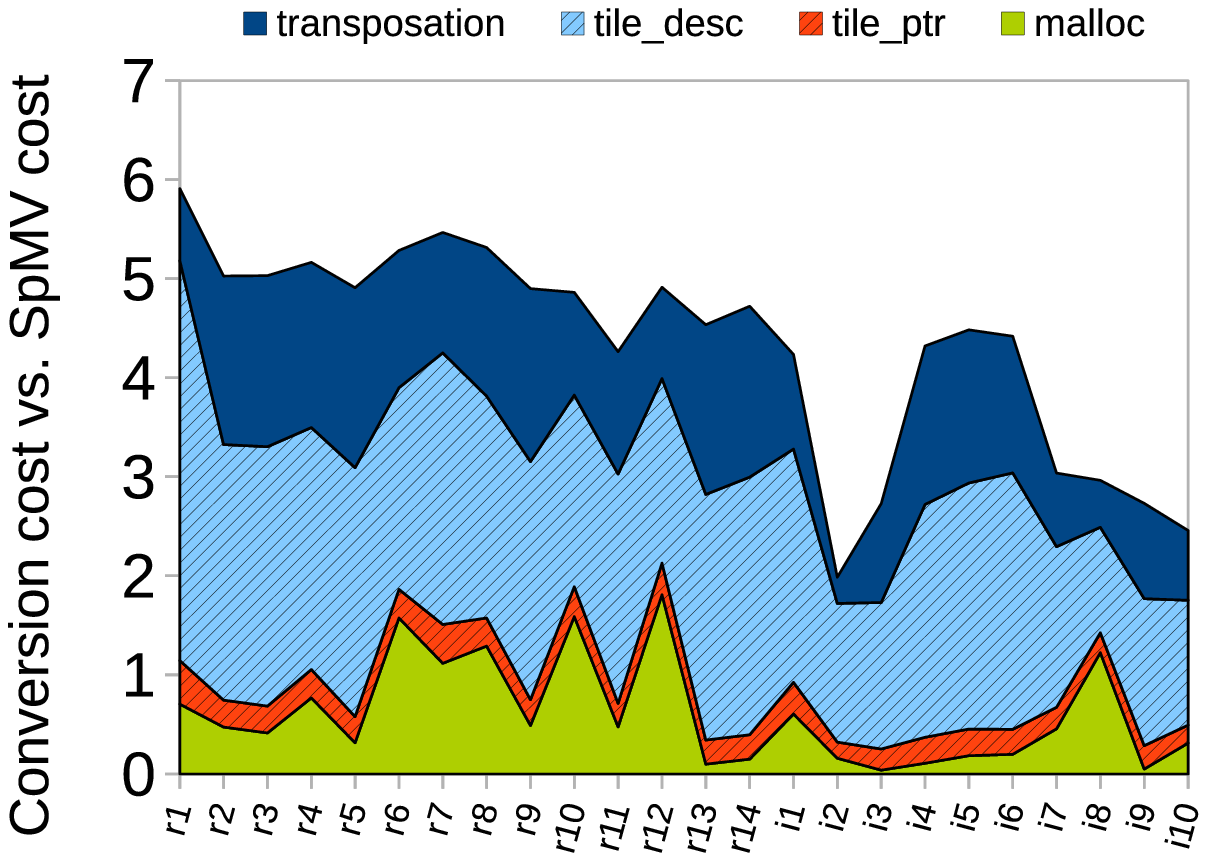, trim=0in 0in 0in 0.3in, clip=true, width=1.6in}}
\subfloat[(d) The Xeon Phi.]{\epsfig{file=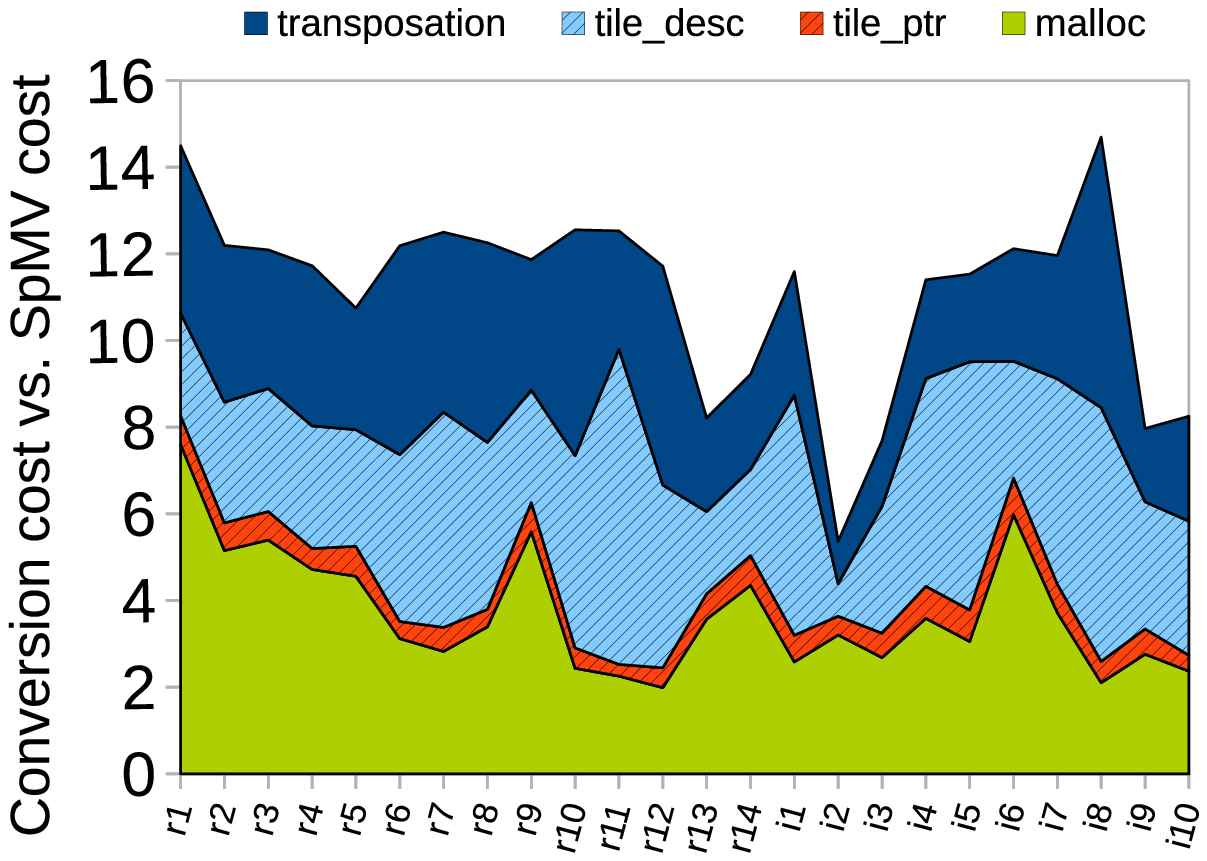, trim=0in 0in 0in 0.3in, clip=true, width=1.6in}}
\caption{The normalized format conversion cost.}
\label{fig.conv}
\end{figure}

The format conversion from the CSR to the CSR5 includes four steps: (1) memory allocation, (2) generating \texttt{tile\_ptr}, (3) generating \texttt{tile\_desc}, and (4) transposition of \texttt{col\_idx} and \texttt{val} arrays. Figure~\ref{fig.conv} shows the cost of the four steps for the 24 matrices (the x axis is the matrix \textit{id}s) on the four used platforms. Cost of one single SpMV operation is used for normalizing format conversion cost on each platform. We can see that the conversion cost can be on average as low as the overhead of a few SpMV operations on the two GPUs. On the two x86 platforms, the conversion time is longer (up to cost of around 10--20 SpMV operations). The reason is that the conversion code is manually SIMDized using CUDA or OpenCL on GPUs, but only auto-parallelized by OpenMP on x86 processors. 

\subsection{Iteration-Based Scenarios}

Since both the preprocessing (i.e., format conversion from a basic format) time and the SpMV time are important for real-world applications, we have designed an iteration-based benchmark. This benchmark measures the overall performance of a solver with $n$ iterations. We assume the input matrix is already stored in the CSR format. So the overall cost of using the CSR format for the scenarios is $nT^{csr}_{spmv}$, where $T^{csr}_{spmv}$ is execution time of one CSR-based SpMV operation. For a new format, the overall cost is $T^{new}_{pre}+nT^{new}_{spmv}$, where $T^{new}_{pre}$ is preprocessing time and the $T^{new}_{spmv}$ is one SpMV time using the new format. Thus we can calculate speedup of a new format over the CSR format in the scenarios, through $(nT^{csr}_{spmv}) / (T^{new}_{pre}+nT^{new}_{spmv})$.

Table~\ref{tab.iter} shows the new formats' preprocessing cost (i.e., $T^{new}_{pre}/T^{new}_{spmv}$) and their speedups over the CSR format in the iteration-based scenarios when $n=50$ and $n=500$. The emboldened font in the table shows the highest positive speedups on each platform. The compared baseline is the fastest CSR-based SpMV implementation (i.e., Intel MKL, nVidia cuSPARSE/CUSP, CSR-vector from CUSP, and Intel MKL, respectively) on each platform. We can see that because of the very low preprocessing overhead, the CSR5 can further outperform the previous methods when doing 50 iterations and 500 iterations. Although two GPU methods, the ACSR format and the CSR-Adaptive approach, in general have shorter preprocessing time, they suffer from lower SpMV performance and thus cannot obtain the best speedups. On all platforms, the CSR5 always achieves the highest overall speedups. Moreover, the CSR5 is the only format that obtains higher performance than the CSR format when only 50 iterations are required.

\section{Related Work}

A great deal of work has been published on accelerating the SpMV operation. The \textbf{block-based sparse matrix construction} has received most attention~\cite{Ashari:An, Buluc:Reduced, Buluc:Parallel, Choi:Model, Pinar:Improving, Vuduc:OSKI, Yan:yaSpMV} because of two main reasons: (1) sparse matrices generated by some real-world problems (e.g., finite element discretization) naturally have the block sub-structures, and (2) off-chip load operations may be decreased by using the block indices instead of the entry indices. However, for many matrices that do not exhibit a natural block structure, trying to extract the block information is time consuming and has limited effects. 

On the other hand, the \textbf{hybrid formats}~\cite{Bell:Implementing, Su:clSpMV}, such as HYB, have been designed for irregular matrices. However, higher kernel launch overhead and invalidated cache among kernel launches tend to decrease their overall performance. Moreover, it is hard to guarantee that every sub-matrix can saturate the whole device. In addition, some relatively simple operations such as solving triangular systems become complex while the input matrix is stored in two or more separate parts.

The recent \textbf{row block methods} showed good performance either for regular matrices~\cite{Greathouse:Efficient} or for irregular matrices~\cite{Ashari:Fast}, but not for both. In contrast, the CSR5 can deliver higher throughput both for regular matrices and for irregular matrices. 

The \textbf{segmented sum methods} have been used in two recently published papers~\cite{Tang:Optimizing, Yan:yaSpMV} for the SpMV on either GPUs or Xeon Phi. However, both of them need to store the matrix in COO-like formats to utilize the segmented sum. In contrast, the CSR5 format saves useful row index information in a compact way, and thus can be more efficient both for the format conversion and for the SpMV operation.

Sedaghati et al.~\cite{Sedaghati:Automatic} constructed machine learning classifiers for \textbf{automatic selection of the best format} for a given sparse matrix on a target GPU. The CSR5 format described in this work can further simplify such a selection process because it is insensitive to the sparsity structure of the input sparse matrix.

Moreover, to the best of our knowledge, the CSR5 is the only format that supports high throughput \textbf{cross-platform SpMV} on CPUs, nVidia GPUs, AMD GPUs and Xeon Phi at the same time. This advantage may simplify the development of scientific software for processors with massive on-chip parallelism.

\section{Conclusions}
In this paper, we proposed the CSR5 format for efficient cross-platform SpMV on CPUs, GPUs and Xeon Phi. The format conversion from the CSR to the CSR5 was very fast because of the format's insensitivity to sparsity structure of the input matrix. The CSR5-based SpMV was implemented by a redesigned segmented sum algorithm with higher SIMD utilization compared to the classic methods. The experimental results showed that the CSR5 delivered high throughput both in the isolated SpMV tests and in the iteration-based scenarios.

\section{Acknowledgments}

The authors would like to thank James Avery~(KU), Huamin Ren~(AAU), Wenliang Wang~(BOC), Jianbin Fang~(NUDT), Joseph L. Greathouse~(AMD), Shuai Che (AMD), Ruipeng Li~(UMN), Anders Logg~(Chalmers and GU), and our anonymous reviewers for their insightful feedback. We thank Klaus Birkelund Jensen~(KU), Hans Henrik Happe~(KU) and Rune Kildetoft~(KU) for access to the Intel Xeon and Xeon Phi machines. We thank Bo Shen~(Inspur) for helpful discussion about Xeon Phi programming. We also thank Arash Ashari~(OSU), Intel MKL team, Joseph L. Greathouse~(AMD) and Mayank Daga~(AMD) for sharing source code, libraries or implementation details of their SpMV algorithms with us. Finally, we thank the PPoPP '15 reviewers for their valuable suggestions and comments.


%
%
%

\end{document}